\documentclass[useAMS,usenatbib]{mn2e}
\usepackage{graphicx}
\usepackage{natbib}
\usepackage{ulem}

\title[The Luminosity Function of the NoSOCS Galaxy Cluster Sample]{The Luminosity Function of the NoSOCS Galaxy Cluster Sample}
\author[E. De Filippis et al.]
  {E.~De Filippis,$^1$\thanks{E-mail:bettydefilippis@gmail.com}
  M.~Paolillo,$^{1,2}$\thanks{E-mail:paolillo@na.infn.it} G.~Longo,$^{1,2,3}$ F.~La Barbera,$^4$ R.R.~de Carvalho,$^5$ R.~Gal$^6$
\\
  $^1$Dip. di Scienze Fisiche, Universit\`a di Napoli "Federico II," Compl. Univ. di Monte S. Angelo V. Cinthia, 9, I-80126, Napoli, Italy\\
  $^2$ Istituto Nazionale di Fisica Nucleare, Sez.di Napoli, Italy\\
   $^3$ Department of Astronomy, California Institute of Technology, Pasadena, USA\\
  $^4$ INAF-OACN, via Moiariello 16, I-80128 Napoli, Italy\\
  $^5$ Instituto Nacional de Pesquisas Espaciais - Divis\~{a}o de Astrof\'{i}sica(CEA), S\~{a}o Jos\'{e} dos Campos, SP 12227-010, Brazil\\
  $^6$ University of Hawai'i, Institute for Astronomy, 2680 Woodlawn Dr., Honolulu, HI 96815, USA\\
}

\begin{document}

\date{}

\pagerange{\pageref{firstpage}--\pageref{lastpage}} \pubyear{2002}

\maketitle

\label{firstpage}

\begin{abstract}
We present the analysis of the luminosity function of a large sample of galaxy
clusters from the Northern Sky Optical Cluster Survey, using latest 
data from the Sloan Digital Sky Survey. Our global luminosity function (down to $M_r\lse -16$) does not show the presence of an ``upturn'' at faint magnitudes, while we do observe a strong dependence of its 
shape on both richness and cluster-centric radius, with a
brightening of M$^*$ and an increase of the dwarf to giant ratio with richness, 
indicating that more massive systems are more
efficient in creating/retaining a population of dwarf satellites.
This is observed both within physical ($0.5\ {\rm R}_{200}$)
and fixed ($0.5\ {\rm Mpc}$) apertures, suggesting that the trend is either
due to a global effect, operating at all scales, or to a local one but operating on even smaller scales. We further observe a decrease of the relative number of dwarf galaxies
towards the cluster center; this is most probably due to tidal collisions or collisional disruption of the dwarfs since merging processes are inhibited by the high velocity dispersions in cluster cores and, furthermore, we do not observe a strong
dependence of the bright end on the environment.

We find indication that the dwarf to giant ratio decreases with increasing redshift, within $0.07\leq z<0.2$. We also measure a trend for stronger suppression of faint galaxies (below M$^*+2$) with increasing redshift in poor systems, with respect to
more massive ones, indicating that the evolutionary stage of less
massive galaxies depends more critically on the environment.  

Finally we point out that the luminosity function  is far from universal; hence the uncertainties introduced by the different methods used to build 
a composite function may partially explain the variety of
faint-end slopes reported in the literature as well as, in some cases,
the presence of a faint-end upturn. 
\end{abstract}


\section{Introduction}
Due to its integrated nature, the luminosity function (LF) of a specific class of objects can be used to study the distribution of luminous matter in the Universe, after taking into account the systematic uncertainties due to  cosmic variance \citep[e.g.][]{bin88,bla01,robe10}. In particular, the main focus of recent work has been the connection between galaxies and dark matter halos - constraining various physical mechanisms governing the formation and evolution of galaxies (e.g. gas cooling, star formation, etc.). The study of the halo occupation distribution linked to the LF became a key factor in not only understanding physical processes shaping galaxies \citep[e.g.][]{pea00,ber02,bul02,scr02} but also in providing constraints on cosmological models \citep[e.g.][]{zhe07}.
 
Despite  the apparent  simplicity of  deriving  the LF  of galaxy clusters,  and the many works  published in the  last  few  years  \citep{lin96,DeP03,And05,pop05,han05,Gon05,Zan06,Han09}, our ability to  properly characterize  the luminosity  distribution of  these  systems has been  hampered by the  need to establish well  defined, statistically large and robust  samples, and to properly combine  them  to address  an  intrinsically  multiparametric  problem.  In  fact,  the dependence  of  the   slope  of  faint-end  of  the   LF  (i.e.  the giant-to-dwarf  galaxy  ratio)  on  environmental  and  evolutionary parameters is still  debated, as is the presence  of an upturn at faint magnitudes \citep{han05,zuc09}.

In the past, LF studies were based on selection in a single waveband, but today this has changed dramatically with surveys spanning from the ultraviolet to the infrared and radio bands. In particular, in the optical,
large photometric surveys are now available allowing the use of cluster richness \citep[a proxy for mass, see for instance][]{hil10,man10} to characterize galaxy systems. Richness is often measured as the number of galaxies within a given luminosity range and
within a certain distance from the cluster center \citep[e.g.][]{dal92,pos96,gal03} allowing to stack system in richness bins and measure the LF over a wide range of host halo masses (see \citet{gla05} for different richness definition). The purpose of the present  work is to  approach the  problem analyzing the  uncertainties introduced by different reduction and analysis techniques adopted in the literature
using a large sample of galaxy clusters with well defined photometry.
Thus we can establish  a  firm basis for determining which results  are robust  and which depend on the specific choices for cluster detection made by different authors \citep[e.g.][]{ols99,pos02,gal09}.\\
This paper is organized as follows: Section \ref{sec:data} describes the cluster
catalog we use, discussing the cluster properties as well. In Section
\ref{sec:background} the statistical background subtraction is discussed in detail as
this is one of the main components required to estimate individual
cluster LFs, which are presented in Section \ref{sec:LF}. Section \ref{sec:combineLF} presents the
methods used in this paper for composing the individual LFs, either
through a non-parametric or parametric approach. In Section \ref{sec:res_disc}, we
discuss the main results obtained here, including the dependence of
the LF on environment and its redshift evolution. 
Conclusions are drawn in Section \ref{sec:conclusions}.\\
Throughout this  paper we  use a cosmology with ${\rm H}_0=70\  {\rm km\  s}^{-1}\ {\rm Mpc}^{-1}$ ($\Omega_m=0.3$, $\Omega_{\Lambda}=0.7$).


\section{Data}
\label{sec:data}
\subsection{Cluster Catalog}
\label{sec:cl_catalogue}
The Northern Sky Optical Cluster Survey (hereafter NoSOCS,
\citealt{gal09}) is a new, objectively defined catalog of galaxy
clusters drawn from the Digitized Second Palomar Observatory Sky
Survey (DPOSS).  Clusters are detected in two steps. First, the
positions of galaxies from DPOSS are used to generate adaptive kernel
density maps.  Then, S-Extractor is run to detect peaks in the density
maps, which are identified as cluster candidates. 
A detailed
description of the survey and cluster detection technique can be found
in~\cite{gal00,gal03}. Details of the photometric calibration and
star/galaxy separation are discussed in \citet{gal04} and \citet{Ode04}.
The original catalog has been recently updated by improving the definition of bad areas, masking out very bright objects on the original DPOSS data, 
and by performing photometric redshift and cluster richness ($N_{gal}$) estimates for all detected clusters \citep{gal09}.
Enhancement in the photometric redshift measurements
based on DPOSS photometry involve measurement of the fore- and background
galaxy contamination in the cluster area.
The 3-clipped medians of the color and magnitude distributions from
ten background regions are used as the background
correction for that cluster. The redshift estimator is run ten times for each cluster candidate. During the photometric redshift measurement process,
the cluster positions are also recomputed, leading to a $\sim 30\%$ improvement over the $z_{phot} \approx 0.033$ of \cite{gal03}.
At each iteration of the zphot computation, the median
position of the galaxies within a $1\ {\rm h}^{-1}$ Mpc radius of the
previously determined center is calculated, and this is taken as the
new cluster centroid for the next iteration of the photometric
redshift estimation.
The resulting final NoSOCS catalog consists of
$15,502$ clusters at redshift z$\leq 0.4$.
The catalog comprises two separated areas of the sky:
i) the North Galactic Pole (NGP) region covering $8494\ {\rm deg}^2$,
and ii) the Southern Galactic Pole (SGP), corresponding to a $2917\
{\rm deg}^2$ sky region.
The total contamination of the NoSOCS cluster sample amounts to about 
$8\%$ \citep[for details see][]{gal09}.
For very rich clusters ($N_{gal} > 50$), contamination is negligible, 
while it rises above $5\%$ for $N_{gal} < 20-25$. In the 
present work, to keep contamination rate below $5\%$, we focus on a 
sub-sample of NoSOCS, selecting only clusters with $N_{gal} > 25$, in the redshift range $0.07 \le z < 0.2$ 
\citep[as outside these limits the sample is poorly defined; see][]{gal09}.

\begin{figure}
\centering
\includegraphics[height=0.5\textwidth,angle=-90]{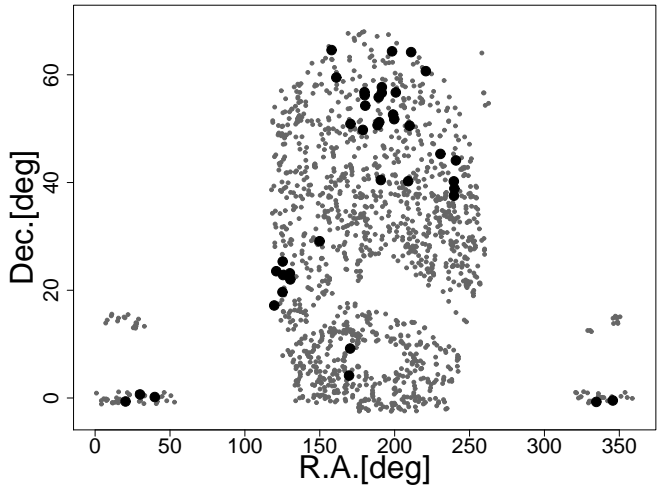}
\caption{Projected distribution  of the sub-sample  of NoSOCS clusters
  analyzed in the present work (gray dots). Black solid circles show the random regions where the global background is computed, to check robustness of the local background determination (see \S~\ref{sec:background}).}
\label{fig:cluster_distribution}
\end{figure}

Since the LFs are computed using galaxy photometry from the
Sloan Digital Sky Survey (SDSS), hence taking advantage of its better
photometric accuracy relative to DPOSS (see below), our analysis is restricted
to clusters found in the area imaged by SDSS.  This is done
by requiring that, for a given cluster, the entire region we use to
derive the LF (see \S~\ref{sec:background}) is enclosed inside the
SDSS area.  These selections result into a final sample of $1,451$
galaxy groups and clusters, whose distribution on the sky is plotted
in Fig.~\ref{fig:cluster_distribution}

\subsection{Galaxy Catalog}
\label{sec:phot_data}

The galaxy catalog is obtained from the Sloan Digital Sky
Survey\footnote{\tt http://www.sdss.org/} 6th data release (hereafter
SDSS-DR6), covering a total sky area of $9583\ {\rm deg}^2$
\citep{fuk96,gun98,yor00,ade08}. For each cluster, we select all
galaxies within a $10\times 10$~${\rm Mpc}^2$ region, centered on the
cluster centroid.  
The SDSS photometry of point-like sources is 95$\%$ complete down to an 
$r'$-band model magnitude of ${\rm m_r}=22$~\citep{sto02}. 
Since the SDSS star/galaxy classification is still reliable down to $m_r \sim 21.5$
(\citealt{lup01}; see also \citealt{cap09}), we select all galaxies down to the 
latter magnitude limit, and adopt this value as
the apparent completeness magnitude $m_c$ of the galaxy catalog of a given
cluster. This choice also ensures that there are no threshold effect when applying the K-corrections discussed below, since the completeness limit of the galaxy catalog is 0.5 mag deeper than our adopted cut.
The completeness absolute magnitude of the NoSOCS sub-sample used in this work,
then ranges from around $M_c=-16$ for the low-redshift clusters ($z \sim
0.07$), to about $M_c=-19$ for the upper redshift limit of $z \sim 0.2$.
We retrieve only galaxies with clean photometry from SDSS, by
selecting only $PRIMARY$ objects with SDSS photometry
flags~\footnote{\tt See also\\
http://cas.sdss.org/astrodr6/en/help/docs/realquery.asp\#errflag} 
set following~\cite{yas01}. 
In order to consider only regions for which the galaxy catalog has homogeneous 
photometric accuracy and completeness characteristics, we  also exclude  
objects within circular  regions  around bright stars and large 
galaxies, from the Tycho-2  and RC3 catalogs~\citep{deV91}, respectively.  
The  radius $r$ of  the masked 
regions  is chosen  following the prescriptions of \cite{gal09}. For 
Tycho-2 stars we set ${\rm r}=2\arcmin$ for ${\rm  m}_{\rm Tycho}<7.0$,  
${\rm r}=1.5\arcmin$  for $7.0  \leq {\rm
  m}_{\rm Tycho}  \leq 8.0$, ${\rm  r}=1.0\arcmin$ for $8.0  \leq {\rm
  m}_{\rm  Tycho} \leq  9.5$, while  for galaxies  in the  RC3 catalog
we adopt ${\rm r}=5 \times{\rm r}_{\rm  RC3}$ for ${\rm r}_{\rm RC3}<25\arcsec$
and  ${\rm r}=8\times {\rm  r}_{\rm RC3}$  for ${\rm  r}_{\rm RC3}\geq
25\arcsec$.

The galaxy LF is measured in the $r'$ band, using model galaxy
magnitudes from the SDSS Photo pipeline~\citep{lup01, sto02}.
Magnitudes are corrected for Galactic extinction according
to~\cite{sch98}.  For all galaxies in the region of a given cluster,
apparent magnitudes are converted to absolute magnitudes by the
relation:
\begin{equation}
M  = m  - 5  \log_{10} (D_L  / 10\  {\rm pc})  - {\rm  K}(z),
\end{equation}
where D$_{\rm L}(z)$ is the luminosity distance, and ${\rm K}(z)$ is
the k-correction.  In order to compute D$_{\rm L}(z)$, we assume all
galaxies in the given region to be at the same redshift as the
cluster. While this is correct for cluster galaxies (considering the
lower redshift limit of the NoSOCS catalog), it is certainly incorrect
for foreground/background galaxies. However, the contribution of field
galaxies is statically removed when computing the LF, making the
computation of D$_{\rm L}(z)$, on average, statistically correct. 

\begin{figure}
\begin{center}
\includegraphics[height=0.45\textwidth,angle=-90]{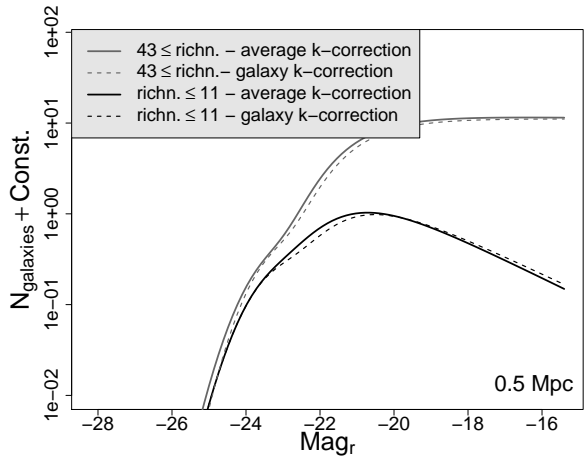}
\caption{Best-fits LFs obtained using the the Maximum-Likelihood 
  approach (Sec.~\ref{sec:ML}) by applying an average k-correction value
  from~\citet{fuk95} (solid line) or k-correcting each  single galaxy
  independently with  the software $kcorrect$ (dashed line). Gray and black colors
  show the cases of rich  and poor clusters, classified according to the richness parameter Richn.$_{\rm SDSS}$ (see Sec.~\ref{sec:cl_pars}).}
\label{fig:compare_KCorrs}
\end{center}
\end{figure}

In order to test the robustness of our results with respect to the method
used to calculate the ${\rm K}(z)$, we adopt two independent
approaches, by (i) using an average k-correction for all galaxies, and
(ii) estimating a specific k-correction for each galaxy.  In case (i),
we adopt the k-correction for elliptical galaxies from~\cite{fuk95} at the cluster redshift.
In case (ii), we estimate the ${\rm K}(z)$ by using the software
$kcorrect$ (version $4\_1\_4$,~\citealt{Bla03b}), 
through a rest-frame filter obtained by blue-shifting
the throughput curve of the SDSS r-band by a factor $(1+z_0)$.  
For $z_0=0$, one recovers
the usual k-correction. As in previous works based on SDSS data
(e.g.~\citealt{Hog04}), we have adopted $z_0=0.1$.  For galaxies at
redshift $z=z_0$, the k-correction is equal to $-2.5 \log (1+z_0)$,
independent of the filter and the galaxy spectral type.  Hence, since
the value of $z_0=0.1$ is very close to the median redshift of the
NoSOCS cluster catalog (see ~\citealt{gal04}), this choice of $z_0$
allows uncertainties on k-corrections to be minimized
(see~\citealt{Bla03b}).  We run $kcorrect$ using the $ugriz$ SDSS
model magnitudes, and the best estimate available for the redshift $z$
of each galaxy, i.e. either the spectroscopic redshift or the
photometric estimate, when the former is not available.  The two
approaches have different advantages and drawbacks.  In case (i), we
are assuming that early-type spectral types dominate the cluster
galaxy population at all magnitudes.  While this is only a rough
approximation, it allows us to avoid bringing the uncertainties on the
k-correction of each single galaxy into the computation of the LF.  On
the other hand, method (ii) implies a larger uncertainty on the
$K(z)$'s, but corrects each galaxy according to its proper spectral
type, inferred from the photometric information. Since the 
foreground/background contaminants are statistically removed from the LF, both
methods should be statistically insensitive to the $K(z)$
of field galaxies. Fig.~\ref{fig:compare_KCorrs} shows the LFs of
galaxies obtained with both methods to estimate the k-correction, for
two richness bins of the parent clusters.  The two methods provide
very similar LFs in both cases. 
In what follows, all results are obtained by applying method (i), but all of our results remain essentially unchanged when using
method (ii).

\subsection{Cluster Properties}
\label{sec:cl_pars}
In order to analyze the environmental dependence of the LF, we derive
it as a function of the richness of the parent clusters and the 
cluster-centric distance, obtained as follows.
The optical richness of a cluster of galaxies, i.e. the number of
galaxies in a given magnitude range within a given physical region of
the cluster, is a proxy for its mass~\citep{Kra04,Gon05,Pop07,hil10,man10}.
As there is no best, objective prescription to measure the richness
parameter, we have performed the analysis by using two different
richness estimates.
\begin{itemize}
\item[i)] For each NoSOCS cluster, \cite{gal09} measured the richness
  parameter, $N_{gal}$, as the background subtracted number of
  galaxies in the cluster, within an aperture of $0.5\ {\rm Mpc}$, in
  the (r-band) magnitude range of $-22$ to $-19$. To take advantage of
  the more accurate SDSS photometry (relative to DPOSS), we
  re-measured cluster richness according to this same definition using
  SDSS photometry.  Hereafter, this updated richness parameter is
  indicated as Richn.$_{\rm SDSS}$\footnote{We note that \cite{gal09}
    actually used an iterative method in order to include
    k-corrections in the richness estimates. This refinement is not
    implemented in our procedure.}. This approach has the advantage of
  being simple and commonly used in the literature, as well as
  providing richness estimates well correlated to cluster mass
  \citep{Lop06}. However, it becomes more and more uncertain for
  poorer systems, as it uses only a limited range of the magnitude
  distribution of cluster galaxies.  Notice also that the luminosity
  range in the definition above extends two magnitudes below
  $M^{\star}$, possibly introducing a spurious dependence of the LF
  faint-end shape on cluster richness.
\item[ii)] To minimize the drawbacks in the definition of Richn.$_{\rm
    SDSS}$, we also define a new richness estimate, Richn.$_{\rm ML}$,
  as the integral of the best-fitting Schechter function to the
  cluster LF in the luminosity range ${\rm M_r} \leq-21.0$. In order
  to account for the possible dependence of the Schechter fit on
  cluster richness (see \S~\ref{sec:richness} ), we estimate
  Richn.$_{\rm ML}$ using an iterative procedure.  First, we split our
  cluster sample according to Richn.$_{\rm SDSS}$, and derive the
  Schechter fit to the LF for each richness bin. For each cluster in
  the bin, the LF fit is re-scaled to match the number counts of
  galaxies brighter than $-21$ .  This provides a first richness
  estimate, Richn.$^0_{\rm ML}$.  We then re-arrange the cluster
  sample based on Richn.$^0_{\rm ML}$, and repeat the procedure,
  obtaining the second, final richness estimate, Richn.$_{\rm ML}$.
  The Schechter function fits are obtained with the Maximum-Likelihood
  approach described in Sec.~\ref{sec:ML}. Notice that this iterative procedure
  is preferable to measuring the Richn.$_{\rm ML}$ from the Schechter
  fit to the LF of single clusters, since results of single fits exhibit
  a large measurement scatter (see \S \ref{sec:res_disc}).
  Richn.$_{\rm ML}$ is designed to probe mainly the LF normalization
  at $M^\star$, i.e. the abundance of giant galaxies in the cluster,
  independent of the LF faint-end slope. 
 On the other hand, it is
  based on the integral of a parametric fit re-scaled to the whole magnitude
  distribution of galaxies in a cluster, hence being virtually less
  affected by Poissonian noise on number counts with respect to richness estimates obtained by the number of galaxies in a given magnitude range.
To further verify the dependence of Richn.$_{\rm ML}$ on the LF faint-end slope, we simulated a set of clusters with Richn.$_{\rm ML}$ ranging from 4 up to 15, i.e. the range covered by our cluster sample, and intrinsic $\alpha=-0.8$. We then recomputed Richn.$_{\rm ML}$, fitting the LF with $\alpha$ forced to be larger/smaller by 0.6 than its best-fit value, i.e. $>1\sigma$ in the large majority of cases of Table \ref{tab:richn_bins}; this would correspond to a situation where a cluster is assigned to a completely erroneous group and thus its Richn.$_{\rm ML}$ is measured using the wrong Schechter model. Even is such extreme scenario the variations in Richn.$_{\rm ML}$ are less than 16\% in all cases, and anyway always below the poissonian uncertainties affecting richness estimates based only on galaxy counts.
\end{itemize}

\begin{figure}
\centering
\includegraphics[height=0.5\textwidth,angle=-90]{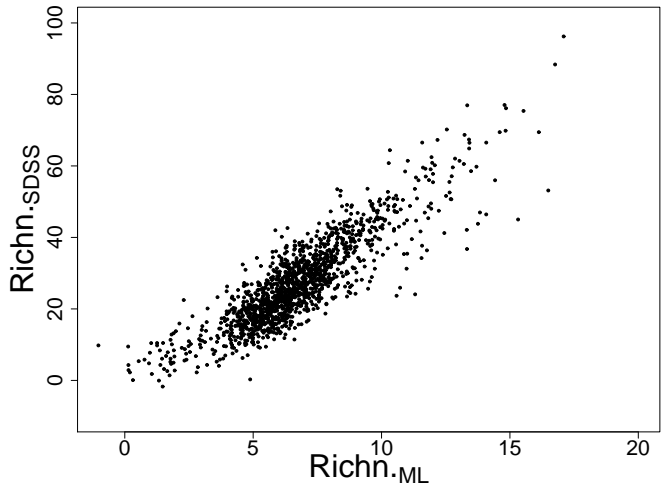}
\caption{Comparison of Richn.$_{\rm SDSS}$ and Richn.$_{\rm ML}$. Notice the good correlation
among the two estimates.}
\label{fig:diff_richnessML}
\end{figure}

Fig.~\ref{fig:diff_richnessML} compares Richn.$_{\rm SDSS}$ with
Richn.$_{\rm ML}$. Despite the different definitions, a good
correlation is observed between the two sets of measurements. More generally, we
verified that our results remain unchanged regardless of which
richness estimate is used.  In particular, the dependence of the LF on
environment is the same using either Richn.$_{\rm ML}$ or
Richn.$_{\rm SDSS}$, even though in the latter case the trends
reported in \S \ref{sec:richness} appear somewhat weaker due to the
reasons discussed above.  For brevity, throughout the rest of the
work, we will only show results obtained for Richn.$_{\rm ML}$.

In order to estimate the cluster-centric distance of galaxies in different
clusters, we use both fixed and, when available, characteristic
radii $R_{200}$.
The characteristic radius of a cluster, R$_{200}$, is the radius
within which the mean inner density is $200$ times the critical
density, $\rho_c(z)$, of the Universe at the cluster redshift. N-body
simulations suggest that the bulk of the virialized mass of a cluster
is generally contained within this radius~\citep[e.g.][]{car97}.
Values of R$_{200}$ were computed for a sub-sample of NoSOCS clusters
from~\cite{gal09}, assuming that the radial distribution of galaxies
within a cluster follows that of the dark matter, and neglecting
possible variations of the mean mass of galaxies with environment.
Of the $1,451$ clusters analyzed in the present work, R$_{200}$
values are available for $814$ clusters. For this sub-sample, the LFs
are derived in
\begin{itemize}
\item three circular regions, with outer radii of $0.2\ {\rm  R}_{200}$, 
 $0.5\ {\rm R}_{200}$ and R$_{200}$, all centered on the cluster centroids; 
 \item an outer annulus, with inner and outer radii of $0.5$ and $1$~$R_{200}$.
\end{itemize}

Deriving the LFs within regions sampling different fractions of a
dynamical radius, such as $R_{200}$, rather than fixed-size apertures,
can actually provide a more physically-meaningful way to compare
galaxy populations at different cluster-centric radii.  However, in
order to fully exploit the extensive statistical power of the
NoSOCS cluster sample, we also perform a fixed-aperture analysis of
the LF using the entire sample of $1,451$
clusters.  This also allows us to perform a more direct comparison
with previous works, where $R_{200}$ measurements were not available.
For the entire sample, we derive the LFs within five fixed-size
apertures:
\begin{itemize} 
\item three circular regions of radii: $0.5,\ 1.5$ and $3.0\ {\rm Mpc}$;
\item two concentric annuli, with $0.5<{\rm r}\leq1.5$ Mpc, and $1.5<{\rm r}\leq3.0$ Mpc.
\end{itemize}

Throughout this paper, results using fixed apertures
always refer to the entire cluster sample, while results within the
physical apertures are obtained using the sub-sample of $814$
clusters.

\section{Background Statistical Subtraction}
\label{sec:background}

The LF of a galaxy cluster is defined as the number of galaxies in the
cluster as a function of luminosity.  The primary difficulty in
measuring the LF is that of assessing the membership of galaxies along
the line of sight. Furthermore, projection effects tend to mimic the presence of a large population of dwarf galaxies, hence producing steep faint slopes~\citep{Val01}.
Ideally, one would need spectroscopic redshifts for
each individual galaxy in the cluster area, but this is unfeasible for a
plethora of reasons. Spectroscopic measurements are extremely time
demanding if possible at all: i.e.  when the sample of clusters is
large, when the faint end of the galaxy population has to be analyzed,
when dealing with high redshift clusters.  Several authors have
attempted to use photometric redshifts to assess cluster membership
but, even though these are effective at reducing the back/foreground
contamination, the latter remains non-negligible and statistical
corrections are still required to remove the contribution of
contaminant sources \citep{Tan05,Tan07,Rud09,cap09}.

\begin{figure}
\centering
\includegraphics[width=0.35\textwidth,angle=-90]{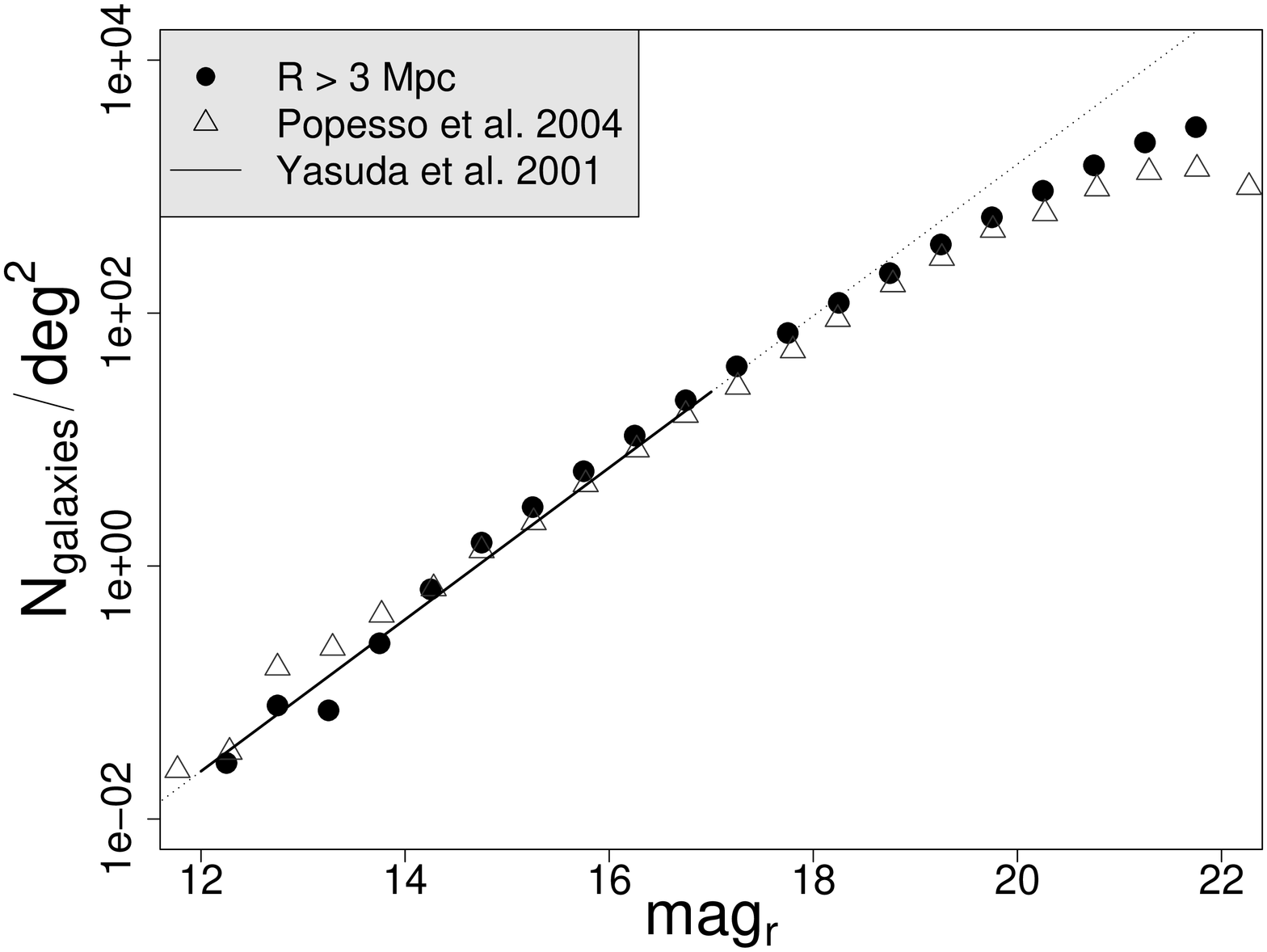}
\includegraphics[width=0.34\textwidth,angle=-90]{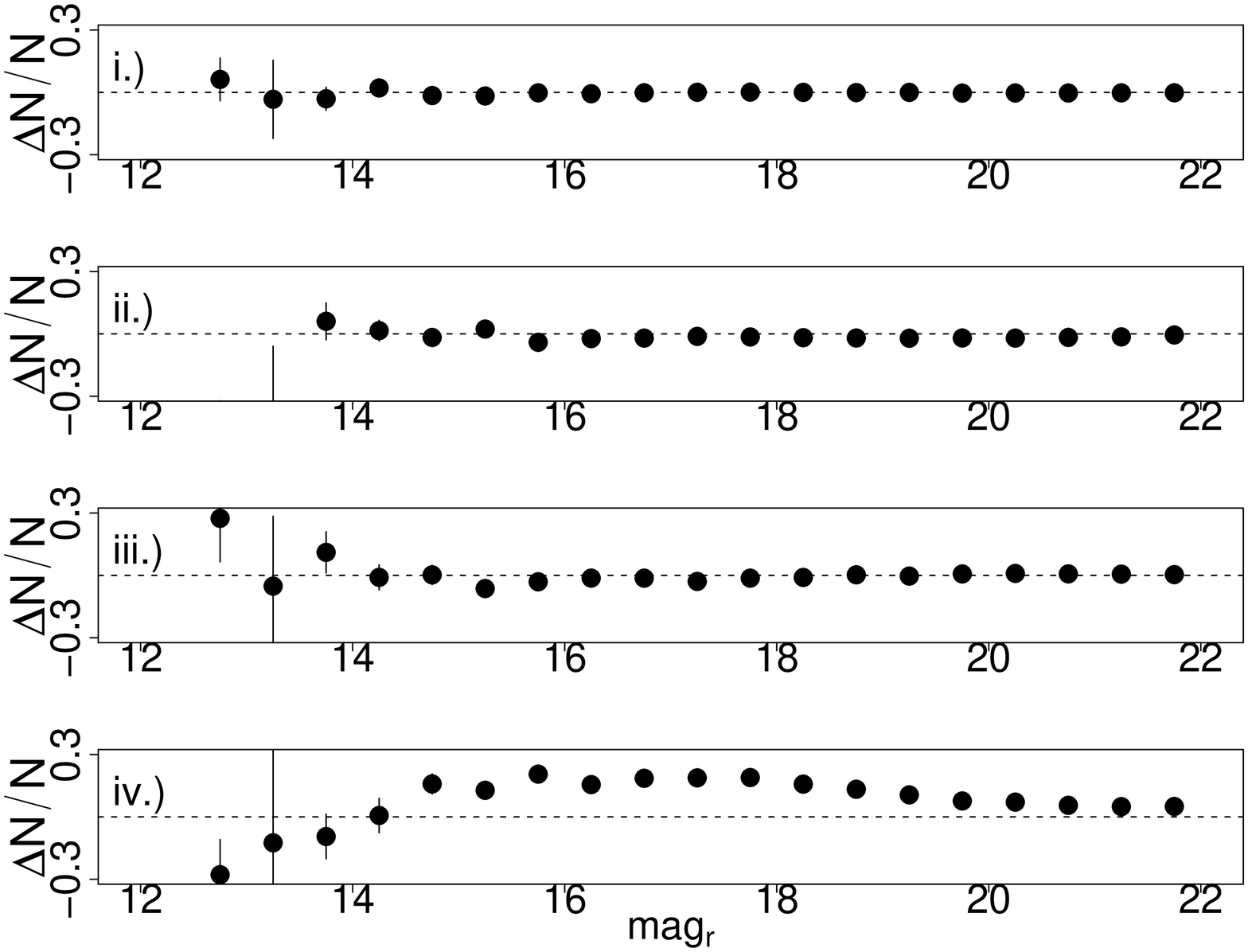}
\caption{Top panel: Average background counts per unit area as a
  function of r-band magnitude (filled circles);  error bars are smaller than symbols.
  Results from \citet{pop04} (empty triangles) are also shown.
  The solid line plots the count-magnitude relation predicted
  from~\citet{yas01} in the magnitude range $12\leq mag_r \leq 17$; 
  the dotted line represents its extrapolation to fainter and brighter magnitudes.
  Bottom panels: percentage difference between local background counts per
  unit area, outside $3\ {\rm Mpc}$, and the: i.) local background
  counts outside $4\ {\rm Mpc}$; ii.) local background counts (outside
  $3\ {\rm Mpc}$) for the 200 poorest clusters in the NoSOCS sample;
  iii.) local background counts (outside $3\ {\rm Mpc}$) for the 200
  richest clusters in the NoSOCS sample. iv.) global background (see
  the text).}
\label{fig:comp_bg}
\end{figure}

The statistical background subtraction is performed by estimating the
contribution of non-cluster members to the number counts of galaxies
in the cluster direction, by measuring the projected number counts of field
galaxies outside the cluster region. Two approaches can be used,
measuring: (i) the ``global'' density of field galaxies on large angular
areas~\citep{gla05,han05}, or (ii) the ``local'' background either in control fields close to the cluster, or in annuli centered
on the cluster centroid~\citep{pao01,got02,pop05}. The differences
between these two methods have been extensively analyzed in the
literature, with most authors finding no significant difference
between them~\citep[e.g.][]{dri98,got02,han05,pop05,bar07}. On the
other hand, as noted by \citet[][also see below]{pao01}, when one wants
to isolate just the main cluster signal without analyzing the structure correlated with the cluster,
the local background approach is preferable, as it takes into account possible background variations
in the cluster region, caused by the large-scale structure within which the
clusters are embedded.

We therefore derive the galaxy LF using a local background approach.
The local background is estimated within a $10\times 10$ Mpc region,
centered on the cluster centroid, outside a radius of $3\ {\rm Mpc}$
(about two times the Abell radius), where the contamination from
cluster galaxies is expected to be negligible. In order to avoid
over-estimating the background level because of the possible presence
of back/foreground galaxy groups, we follow the approach of
\cite{pao01}.  We generate a density map of galaxies in the background
region by convolving the projected distribution of galaxies with a
Gaussian kernel of $\sigma=250\ {\rm kpc}$ in the cluster rest frame
(the typical size of a cluster core). Then, we mask out all density
peaks, above the $3\sigma$ level, from the background region. 
Masked out regions cover, on average, $2-3\%$ of the whole background area
and contain less than $2\%$ of all background galaxies.
Clusters for which the area of the masked regions is larger 
than $10\%$ of the total background one are excluded from our analysis.
 The number counts of galaxies in the remaining region, which we call the
``local background'', is adopted to estimate the expected background
counts in the cluster direction, i.e. one of the regions where the LF
is derived (see Sec.~\ref{sec:cl_pars}). Fig.~\ref{fig:comp_bg} (top
panel) shows the local background counts, per unit area, averaged
among all the clusters in the NoSOCS sample. Good agreement is found
with~\citet{pop04}, who estimated background counts within randomly
selected fields, and with the count-magnitude relation expected for a
homogeneous galaxy distribution in a universe with Euclidean geometry,
as obtained by \cite{yas01}.

Fig.~\ref{fig:comp_bg} also shows some tests we performed to check the
robustness of the local background determination.  The bottom panels
show the fractional difference between averaged local background counts obtained
outside $3\ {\rm Mpc}$ and those measured: i) outside $4\ {\rm Mpc}$; 
ii) outside $3\ {\rm Mpc}$ but only for the $200$ poorest clusters in the NoSOCS sample;
iii) outside $3\ {\rm Mpc}$ but only for the $200$ richest clusters.
Panels i.--iii. show no appreciable difference ($<1\%$, on average) in
number counts in the two background regions, implying that, 
on average, we are not overestimating
the background counts, as might be the case if some residual signal
from the cluster would be still detectable in our ``local'' background.
If present, such effect would indeed be less important at larger cluster-centric
distances (panel i.), and/or produce a fictitious higher background
level for richer clusters (panel iii.), while no measurable difference 
is instead observed.  
\cite{she09}, in their weak lensing analysis, measure the excess number density due to the cluster in the background region to vary between $\sim 1\%$ for poorest groups to $\sim 3.5\%$ for rich clusters. Poorest groups are excluded from our cluster sample, and hence the difference in the excess number densities caused by poor and rich clusters in the surrounding background region should be lower than $\sim 2.5\%$, measured over the whole magnitude range, in rough agreement with what found in this work.

As a further test, we also extracted a ``global''
background from $42$ control fields, selected within the SDSS area,
containing no known galaxy clusters.  The control fields have a
radius of $60 \arcmin$ each, covering a total area of $\approx 130\
{\rm deg}^2$. Their distribution on the sky is shown in
Fig.~\ref{fig:cluster_distribution} (black solid circles).  
The difference between the counts measured in this ``global'' background, 
and those obtained in the ``local`` one (outside $3\ {\rm Mpc}$),
is plotted in panel iv. of Fig.~\ref{fig:comp_bg}. 
While for bright magnitudes, $m_r < 14$, the
two background estimates are consistent within the errors, the global
background tends to be systematically lower than the local one for
fainter magnitudes, in agreement with the findings
of~\cite{pao01}. 
Since we can reasonably exclude that our background local estimate is 
over-estimated because of contamination from cluster galaxies (see panel i. in Fig.~\ref{fig:comp_bg}), we can conclude that the local
background also includes the contribution of the large-scale structure
around clusters and groups of galaxies, which is instead not accounted for by
the number counts in the control fields. Throughout
the present work we then compute the LFs by using the local rather global background
determination. We note that, while many authors have claimed no
significant differences in the LF when using either the local or
global background, Fig. 4 of~\citet{pop04} reveals, in agreement with our findings,
a tendency for the LF to have a shallower faint-end slope when using global background
counts, implying a differential slope between the local and global
background galaxy counts, which could affect the slope of 
the faint-end slope of the resulting LF.

\section{Individual Luminosity Functions}
\label{sec:LF}
In order to derive the LF of galaxies in individual NoSOCS clusters,
for each cluster we estimate the number counts of galaxies in the
local background region (see previous section) and subtract them from
the number counts of galaxies in the cluster region (see
\S~\ref{sec:cl_pars}). Number counts are computed in half-magnitude
bins. Background counts are rescaled to the effective area of the
cluster region, accounting for excised areas due to bright objects in
both background and cluster regions (\S~\ref{sec:phot_data}).
Errors on the individual LFs are measured following~\citet{pao01}.

\begin{figure}
\centering
\includegraphics[width=0.23\textwidth]{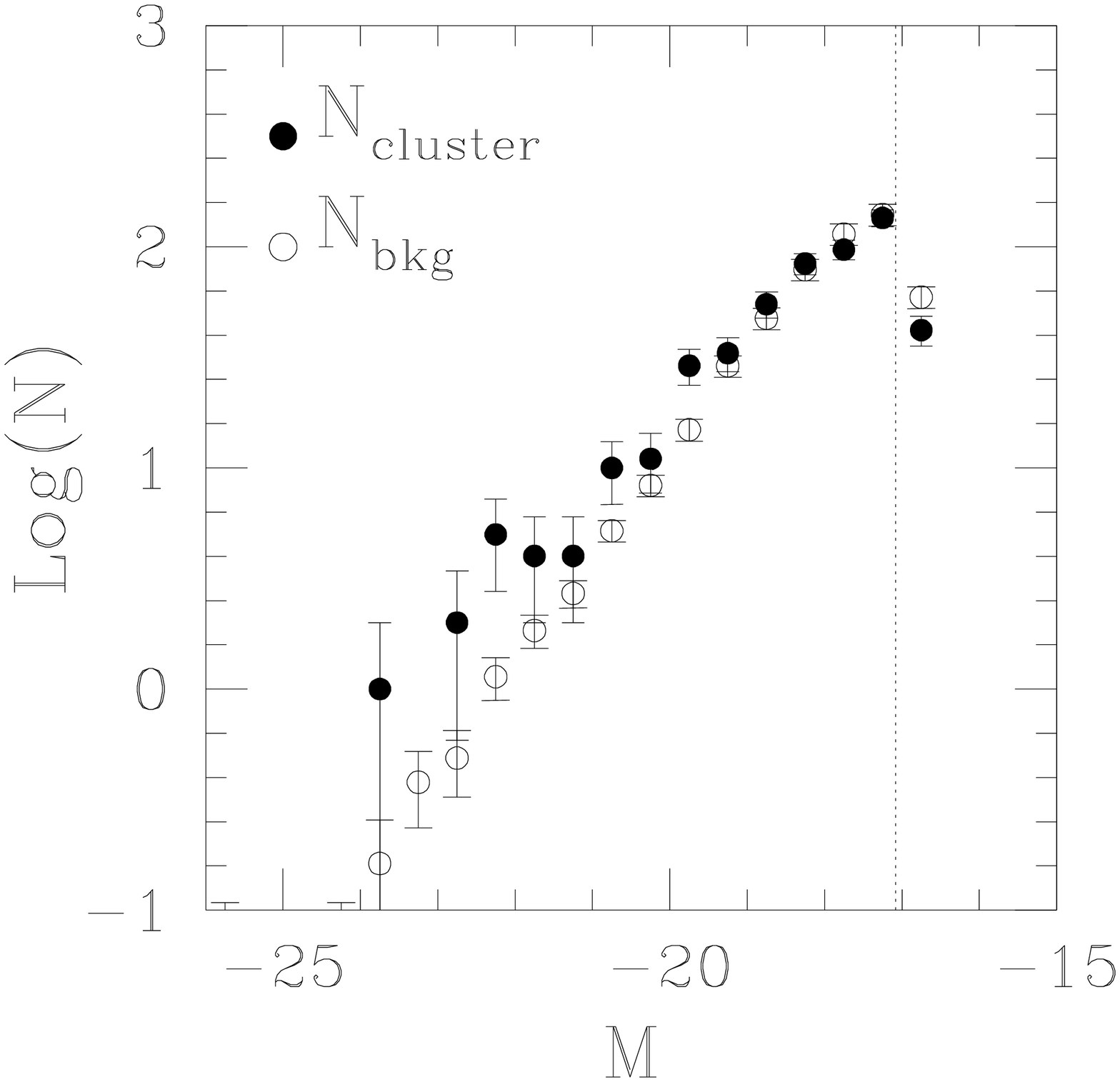}
\includegraphics[width=0.23\textwidth]{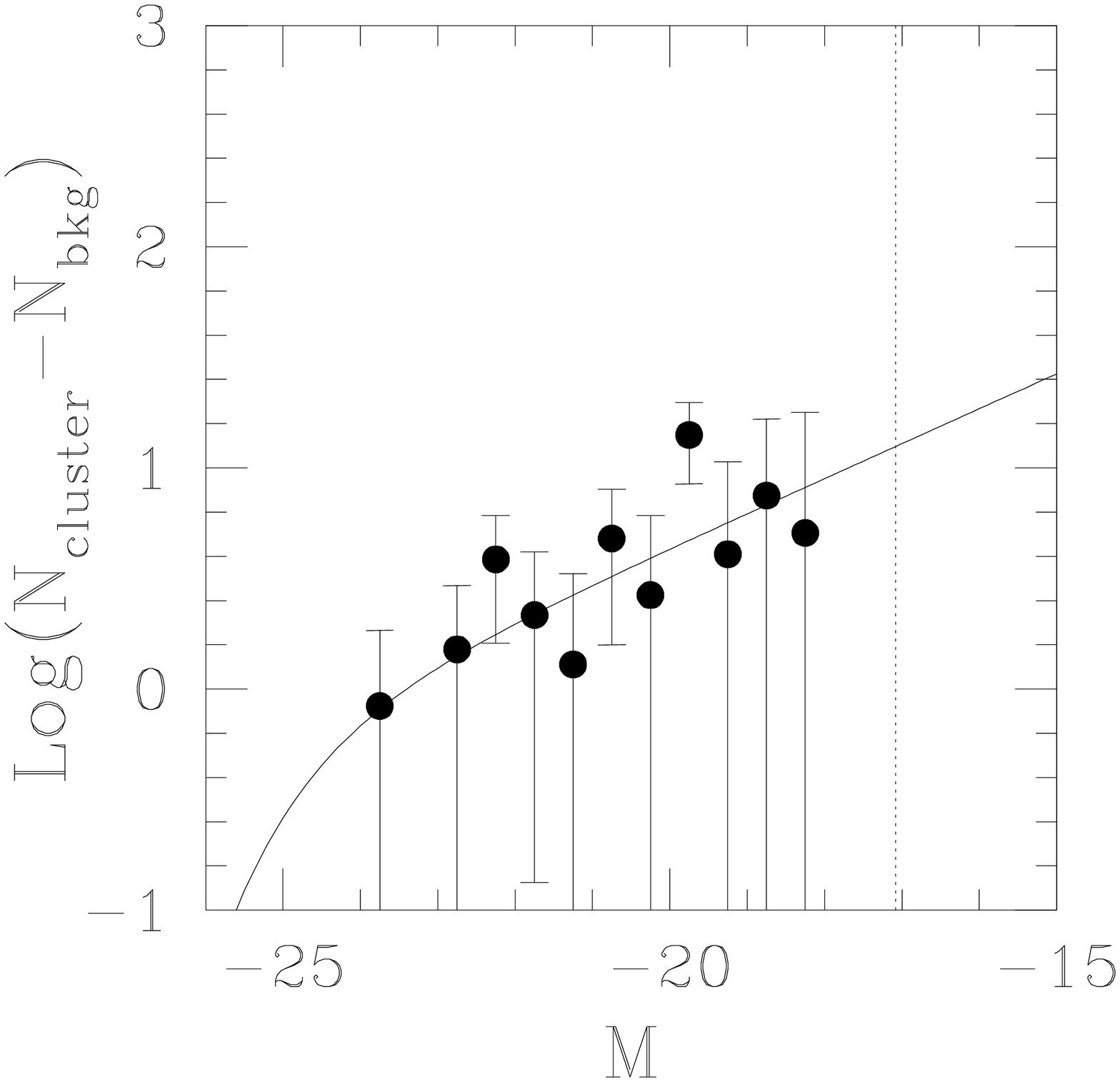}
\includegraphics[width=0.23\textwidth]{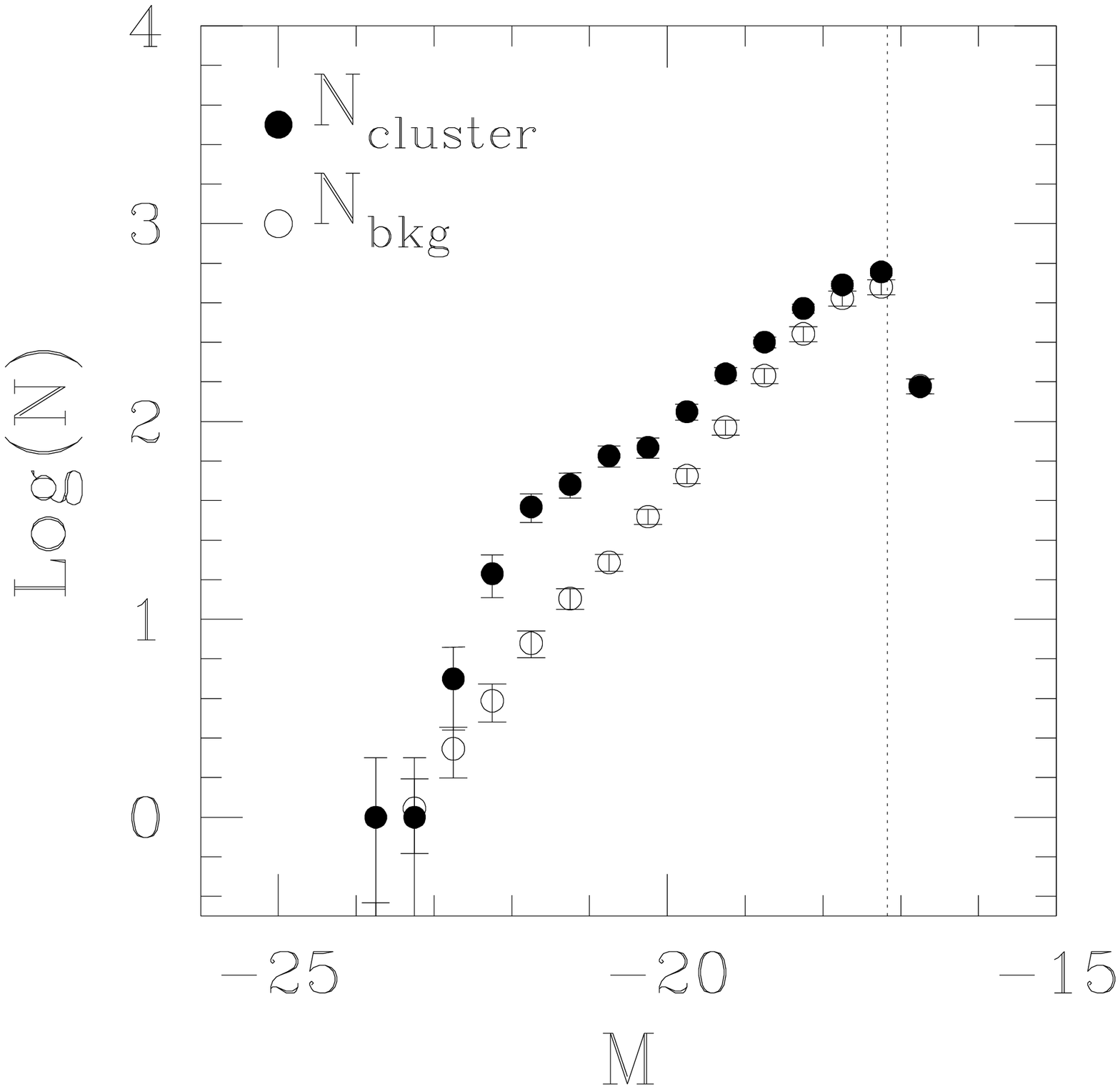}
\includegraphics[width=0.23\textwidth]{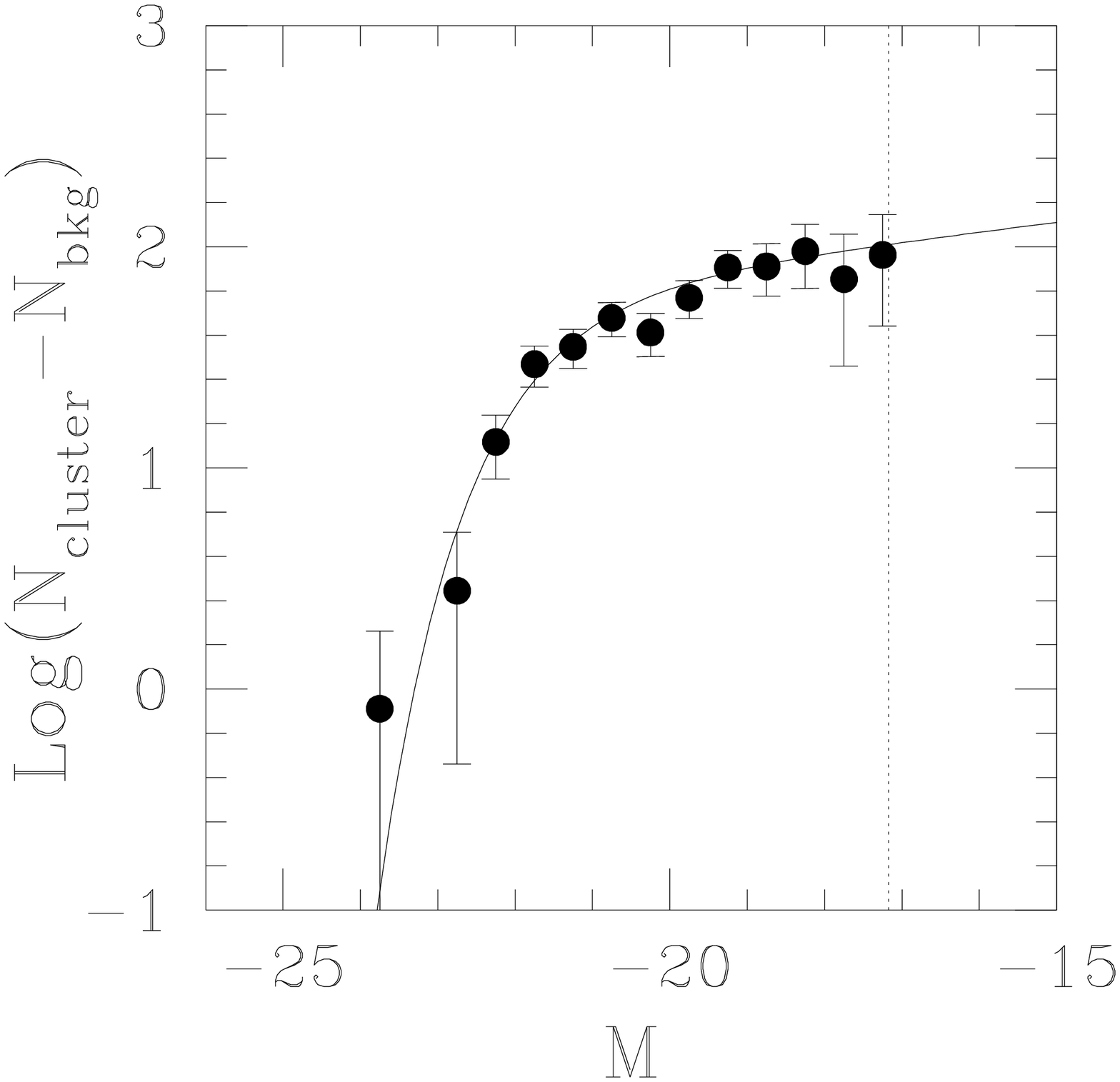}
\caption{Left panels. Galaxy counts extracted within R200 (solid
  circles), and in the local background (empty circles) for the poor cluster 
  NSC14787 (top) and rich cluster NSC09718 (bottom), both located at ${\rm z\approx0.13}$. Right  panels. Background  
  subtracted  number counts, i.e. individual LFs, of the two clusters.  The  
  dashed line  corresponds to the  completeness limit of the SDSS 
  photometry (Sec.~\ref{sec:phot_data}); the solid line shows the best fit models. Poor structures, with low signal LFs, might lead to unrealistic results in the LF best fit (as for NSC14787, top right panel), which cause the large spread in the LF results, shown in Fig.~\ref{fig:single_fit}.}
\label{fig:LF_rich_poor}
\end{figure}

As an example of the procedure to derive the single LFs,
Fig.~\ref{fig:LF_rich_poor} exhibits background and cluster number
counts, as well as the resulting LF for two clusters among the poorest
(NSC14787, ${\rm z=0.132}$, top panels) and richest (NSC09718, ${\rm z=0.137}$,
bottom panels) structures in the NoSOCS sample.
While for rich structures the LF is
significantly detected above the background level, poor structures are
affected by their very low density contrast, as shown by the large
uncertainties on background-subtracted number counts at both the
bright and faint ends of the LF.

In order to characterize the properties of the individual LFs, we fit 
them  by a parametric model given by the Schechter function:
\begin{equation}
\Phi(M)dM=\Phi^{*}\       10^{0.4(M^{*}-M)(\alpha+1)}\exp      \left
(10^{0.4(M^{*}-M)} \right ) dM
\end{equation}
where $M$ is the galaxy magnitude, $\Phi^{*}$ is the normalization
factor, $M^{*}$ is the characteristic knee magnitude and $\alpha$ is
the faint-end slope of the LF.
For each cluster, we fit the background-subtracted counts using a
$\chi^2$ minimization procedure.  To calculate the $\chi^2$, the LF
model is integrated over each magnitude bin.  Uncertainties on the
best-fitting parameters, $\Phi^{*}$, $M^{*}$, and $\alpha$, are
determined by marginalizing each parameter over the remaining ones.
Results are presented in Sec.~\ref{sec:res_disc}.

\section{Composite Luminosity Functions}
\label{sec:combineLF}
To analyze the dependence of the cluster LF on different
properties, such as cluster richness, redshift, and cluster-centric
distance of galaxy populations, we bin the NoSOCS sample with respect
to each quantity, and derive the composite LF of galaxies in each bin.
The composite LFs are derived using two alternative approaches:
by performing a weighted average of the individual cluster LFs (Sec.~\ref{sec:compositeLF}), and by
performing a simultaneous Maximum-Likelihood (hereafter ML) fit to all the
individual cluster LFs in the given bin (Sec.~\ref{sec:ML}). 
The first approach is non-parametric, i.e. we make no prior assumptions on the
shape of the LF, while the ML fit assumes a given functional form.

\subsection{Non-parametric Approach: ``cumulating'' the LF}
\label{sec:compositeLF}
The cumulative LF (hereafter CLF) of a given sample of galaxy clusters
is the weighted mean of the individual LFs.  Previous works have
focused on the differences among alternative cumulation approaches, 
and the following two methods have come out as the most reliable ones:
\begin{description}
\item[{\it Colless method.}]  \citet{col89} derived the CLF as: 
\begin{equation}
N_{\rm cj}=\frac{N_{\rm c0}}{m_{\rm j}} \sum_i \frac{N_{\rm ij}}{N_{\rm i0}},
\label{eq:CLF_COL}
\end{equation}
where $N_{\rm cj}$ is the number of galaxies in the $j$-th bin of the
CLF, ${N_{\rm ij}}$ is the number of galaxies in the $j$-th bin of the
$i$-th cluster LF, $m_j$ is the number of clusters contributing to the 
$j$-th magnitude bin, ${N_{\rm i0}}$ is a normalization factor and:
\begin{equation}
N_{\rm c0}=\sum {N_{\rm i0}}.
\end{equation}
For the $i$-th cluster, the ${N_{\rm i0}}$ is defined as the number of
galaxies brighter than the completeness absolute magnitude, $M_c$,
of the cluster sample. For a flux-limited survey, $M_c$ coincides
with the completeness magnitude of the most distant cluster.  The
error on the resulting CLF is
\begin{equation}
\sigma_{N_{\rm cj}}=\frac {N_{\rm c0}}{m_{\rm j}} \sqrt{\sum_i \left(\frac{\sigma_{N_{\rm ij}}}{N_{\rm i0}}\right)^2},
\end{equation}
where $\sigma_{N_{\rm ij}}$ is the statistical uncertainty on $N_{\rm
  ij}$ (see Sec.~\ref{sec:LF}).  For the NoSOCS sample, we compute the
individual LFs as described in Sec.~\ref{sec:LF}.  The completeness
magnitude of the sample is $M_c=-19$ in the $r'$ band (see
Sec.~\ref{sec:phot_data}), and ${N_{\rm i0}}$ is hence computed as
the field-corrected number of galaxies, of the  $i$-th   cluster,  brighter  than  $-19$. \\
\item[{\it GMA method.}]  \citet{gar99} proposed an alternative
  cumulation method, where one weights each cluster LF according to
  the number of galaxies contained within an adaptive magnitude range.
  In this approach, the CLF is given by:
\begin{equation}
N_{\rm cj}=\frac 1{m'_{\rm j}} \sum_i N_{\rm ij}  w_{\rm i}^{-1}
\label{eq:CLF_GMA}
\end{equation}
where $N_{\rm cj}$ and $N_{\rm ij}$ are defined as in the Colless
method, $m'_{\rm j}$ is the number of clusters with completeness
magnitude fainter than the $j$-th bin, and $w_{\rm i}$ is the
normalization factor of each cluster. For the $i$-th cluster, we denote
as $M_{c,i}$ its completeness (absolute) magnitude. The $w_{\rm i}$ is
given by the ratio of the number of galaxies brighter than $M_{c,i}$
in the cluster, to the average number of galaxies brighter than
$M_{c,i}$ among all clusters whose completeness magnitude goes fainter
than $M_{c,i}$.  The error on the CLF is
\begin{equation}
\sigma_{N_{\rm cj}}=\frac 1{m'_{\rm j}} \sqrt{\sum_i \left(\sigma_{N_{\rm ij}}w_{\rm i}^{-1}\right)^2},
\end{equation}
where $\sigma_{N_{\rm ij}}$ is defined as for the Colless method.
\end{description}

\begin{figure}
\centering
\includegraphics[height=0.45\textwidth,angle=-90]{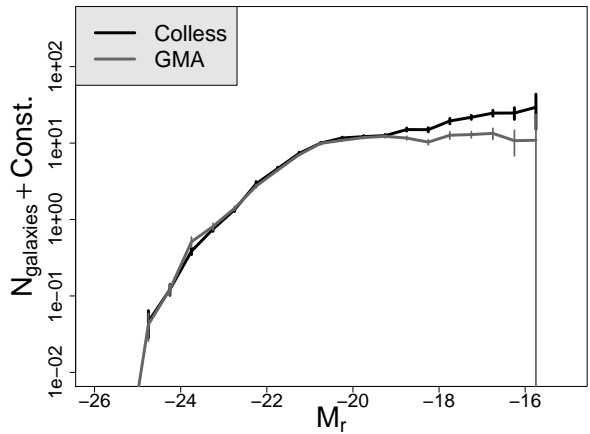}
\caption{Composite LFs (within the cluster region of radius $0.5\ {\rm Mpc}$) 
obtained applying the Colless and GMA  methods plotted, respectively,
 as black and gray solid lines.}
\label{fig:GMA_vs_Colless}
\end{figure}

\begin{figure}
\centering
\includegraphics[width=0.45\columnwidth,angle=-90]{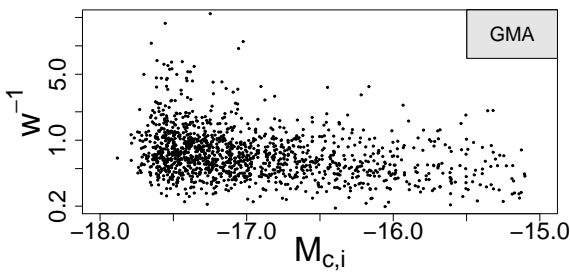}
\caption{Dependence of  the GMA weight, {\it w$_i$}, on  the completeness magnitude, $M_{c,i}$,
  of NoSOCS clusters. Notice that no relevant correlation is detected.}
\label{fig:weights_POP}
\end{figure}

The main difference between the two cumulation methods is that Colless
weights each LF by the number of galaxies in a fixed magnitude range,
while the GMA takes advantage of the different completeness magnitude
of each cluster, hence fully exploiting all the available information
in the data.  In principle, both methods should provide identical
results, at least as far as all clusters exhibit, in a statistical
sense, the same LF, i.e.  the cluster LF is $``universal''$.
Fig.~\ref{fig:GMA_vs_Colless} compares the CLFs obtained by the two
methods for all NoSOCS clusters within a region of
radius $0.5\ {\rm Mpc}$.  The methods agree very well on the bright
end of the LF, while at faint magnitudes, we see a trend for the
Colless method to produce a steeper faint-end than that of the GMA
CLF. A trend in the same direction has been already
reported by~\cite{pop05} (hereafter P05), who detected a dramatic
drop-off in the CLF faint-end when using the GMA, rather than Colless,
method. P05 explained this effect as a result of a strong correlation
among the GMA weights, $w_i$, and the completeness magnitude M$_{\rm
  c,i}$ of the clusters.  This would actually lead to down-weighting the
LFs of clusters having a deeper completeness magnitude 
hence producing a sharp decline in
the CLF faint end.  Fig.~\ref{fig:weights_POP} plots the GMA weights
for the NoSOCS sample as a function of the cluster completeness
magnitude, $M_{c,i}$.  In contrast to P05, we find no strong correlation between
$w_i$ and $M_{c,i}$ (cfr Fig.~5 of P05), implying that the difference
between Colless and GMA CLFs is not caused by the trend found by
P05~\footnote{We notice that P05 define the $w_i$ as the ratio of the
  number of galaxies brighter than $M_{c,i}$ in a cluster, to the {\it
    number}, rather than the {\it average number}, of galaxies
  brighter than $M_{c,i}$. This might explain the discrepancy between
  the lack of trend in Fig.~\ref{fig:weights_POP} of this paper, and the strong trend
  shown in Fig.~5 of P05.}.

The difference between the Colless and GMA methods is instead related to the
fact that the GMA computes the weights by also counting galaxies
in the faint end of the individual LFs. In the case where the
bright part of the LF is identical for all clusters in the sample, all
of the individual LFs have the same weight in the Colless method,
regardless of the shape of the faint-end slope. For the GMA, the
$w_i$'s are the same only if the faint-end part of the LFs is the same
for all clusters. If this is not the case, clusters with a steeper
faint-end have a smaller $w_i^{-1}$, i.e. are down-weighted
in the CLF (see Eq.~\ref{eq:CLF_GMA}).  In other words, if the
individual cluster LFs have different faint-end slopes, the GMA will
give more weight to those clusters with a shallower LF faint-end, hence
producing the difference between the two CLFs, as observed in
Fig.~\ref{fig:GMA_vs_Colless}.  The variation of the
faint-end slope among clusters can be either (i) intrinsic, i.e. the
LF is not universal, or (ii) statistical, because of the
non-Poissonian background fluctuation on the cluster angular scale
(see Sec.~\ref{sec:background}). Poissonian uncertainties
(on both cluster and field counts) do not alter the shape of the
individual cluster LF, as the errors on different magnitudes are not
correlated. This is not the case for the non-Poissonian contribution
to the error budget. As shown in Sec.~\ref{sec:res_disc}, point (i) is
certainly important in driving the difference of the CLFs from the two
methods in Fig.~\ref{fig:GMA_vs_Colless}, as the LF faint-end slope
turns out to depend significantly on cluster richness.

While both the GMA and the Colless methods have the advantage of being
non-parametric, they are also significantly affected by statistical
fluctuations in the individual LFs, in particular when the individual
LFs have low S/N ratio, i.e. for poor groups and for the outskirt
regions of clusters, as well as at the extreme faint end of the LF. At faint
magnitudes, statistical fluctuations in the background signal can even
lead to negative values of field-corrected number counts of the
individual LFs, possibly making the weighted mean in
Eq.~\ref{eq:CLF_COL} and Eq.~\ref{eq:CLF_GMA} ill-defined.  The GMA
method is more sensitive to this issue than the Colless one, as the
weight for each LF is computed by including also the faint range of
the LF, where background fluctuations are more important. 
Therefore, while both cumulation methods
can be applied straightforwardly to well controlled samples of rich structures, 
caution should be taken when analyzing individual LFs with a variety of
S/N ratios and completeness limits.

\subsection{Parametric Approach: the Maximum Likelihood Technique}
\label{sec:ML}
An alternative approach to derive the composite LF of a cluster sample
is to perform a simultaneous Maximum Likelihood fit of
number counts for all clusters. While this method has the drawback of
being parametric, in that a given analytic functional form of the
global LF has to be assumed, it has several advantages relative to the
cumulative approach. First, the data are not binned. Second, no correction has
to be applied for incompleteness at the faint end, and third, it
is not affected by the issue of negative values of field-subtracted
number counts.

The ML approach and its advantages have been thoroughly described
by~\citet{And05}; henceforth we shortly describe only its
implementation for the analysis of the NoSOCS sample. First, we have
to adopt a given functional form for the global LF.  As noted in
Sec.~\ref{sec:LF}, a single Schechter function provides a reasonable
tool to analyze the LF of individual clusters, where the uncertainties
on number counts usually do not allow a detailed analysis of the shape
of the LF.
Now instead we consider a Schechter plus a Log-normal function. The latter term is
used to describe the LF of Brightest Cluster galaxies (BCGs), which
are known not to follow the Schechter distribution typical of faint
galaxies~\citep{Tho93,Biv95,Han09}.

To perform the ML fits, we assume that each galaxy is extracted from a
probability distribution consisting of the above model plus a second
order power-law, representing the background component. For each
cluster, we first fit the second order power-law to the local
background.  The best fit is rescaled to the angular area of the
cluster region. Galaxy counts in all the cluster regions are then
fitted simultaneously, by keeping fixed, for each cluster, its
rescaled background power-law component~\footnote{We tested the
  possibility of fitting simultaneously the local background with the
  number counts in all the cluster regions, but the results turned out
  to be indistinguishable from the case where the local background is
  fitted independently for each cluster.}. The fitting parameters are
the characteristic magnitude, $M^\star$ and the faint-end slope,
$\alpha$, of the Schechter function, the central magnitude,
$M^\star_{BCG}$, and width, $\sigma_{BCG}$, of the BCG's component,
and the two normalization factors of the Schechter and log-normal
functions.

The normalization factors are let free to vary from cluster to
cluster, while the other parameters are set to have the same values
for all clusters. The ML fits are performed using the L-BFGS algorithm
as for the fitting of the individual LFs (Sec.~\ref{sec:LF}).  L-BFGS
is well suited for optimization problems with a large number of
dimensions, as is the case for the CLF fitting, because it never
explicitly forms or stores the Hessian matrix~\citep{Lu94}, still
allowing upper and lower constraints for each variable to be measured.
We also constrain the fitting parameters of the log-normal component of
the CLF to be in the range of $-23.5 \leq M^\star_{BCG}\leq -22.5 $
and $0.3 \leq \sigma_{BCG}\leq 1.0$, based on the typical range
spanned by BCGs \citep[e.g. cf][]{Han09}.  Confidence limits are evaluated by marginalizing over
all unwanted free parameters.

\section{Results and Discussion}
\label{sec:res_disc}
The measurement of the LF of galaxy clusters is an intrinsically
multiparametric problem.  Local and global galaxy
density, density of the intra-cluster medium, age etc. are all known
to affect the relative balance between galaxy populations in clusters
through several effects such as merging, tidal stripping, harassment,
ram pressure stripping and strangulation, whose interplay is not yet well
understood.

We have underlined in the previous sections that the assumption of a
universal LF shape in samples spanning a large range of parameters and
S/N, is not merely an approximation which hides the details of the
processes at work, but may produce biases in the final results
altering the measured LF parameters depending on the technique used to
weight the individual LFs.

\begin{figure}
\centering
\includegraphics[height=0.45\textwidth,angle=-90]{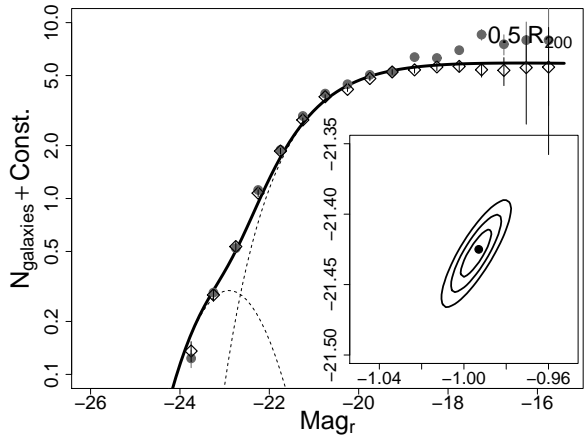}
\caption{ML fit obtained using a Schechter plus a Log-normal components (solid line). The individual components are shown as dashed lines. Data were extracted within $0.5\ {\rm R}_{200}$. Confidence levels for the free parameters of the Schechter function are also plotted. Gray circles and empty diamonds represent the composite LFs obtained using the Colless and GMA methods, respectively.}
\label{fig:LF_band}
\end{figure}

\begin{table}
\centering
\caption{Results ML fit within $0.5\ {\rm R}_{200}$.}
\label{tab:LF_band}
\begin{tabular}{cc}
\hline 
$\alpha$ & $M^\star$\\
\hline
\\
$-0.99_{-0.02}^{+0.01}$	& $-21.42_{-0.04}^{+0.03}$\\
\\
\hline
\end{tabular}
\end{table}

Nevertheless, in order to allow a comparison with previous studies of
large cluster samples, we derived a ``global'' LF of the overall sample within $0.5\
{\rm R}_{200}$ (see Fig.\ref{fig:LF_band} and Table
~\ref{tab:LF_band}).
Differences in cumulation techniques, extraction radii and richness of
the samples affect the shape of the LF (see $\S$~\ref{sec:compositeLF}
and \S~\ref{sec:richness}), making a comparison between independent works
a difficult task. 

In general, the $M^\star$ values agree very well
within the uncertainties among different authors, while the faint-end
slope of the LF is more debated.
Once the different pass-bands and cosmologies are taken into account,
our $M^\star$ is indeed in fair agreement with
\citet{Cra09,gar99,pao01}, with~\cite{Yang09} for galaxy groups in the
SDSS, and with~\cite{Rud09} for red-sequence galaxies, which dominate
the bright end of the LF.

For the steepness of the faint end, our results are in good
agreement with the X-ray selected
clusters by~\citet{Val04} ($\alpha=0.9\pm0.1$), and in marginal agreement
with~\citet{pao01} ($\alpha=1.11^{+0.09}_{-0.07}$). The slightly steeper faint end measured
by~\citet{pao01} is probably an effect of both the larger extraction
radius (${\rm r} \sim 2\ {\rm Mpc}$) and the sample of rich Abell
clusters used. Trends for flatter faint end slopes are instead
observed by~\citet{gar99,Cra09} ($\alpha=-0.81^{+0.05}_{-0.10}$ and $\alpha=-0.84\pm0.32$, respectively). While the average value of $\alpha$
measured by~\citet{Cra09} over the whole low-redshift cluster sample is consistent
with our findings within the uncertainties, the difference
with~\citet{gar99} can be attributed to their smaller extraction radius
($<{\rm r}> \sim 250\ {\rm kpc}$).

\cite{han05,Han09}, fitting a MaxBCG selected sample down to $M_r<-19$,
derive a shallower faint-end. The latter sample however may be skewed
toward poorer systems with respect to NoSOCS, since when
differentiating according to cluster richness their results are in
much better agreement with ours (see \ref{sec:richness}). 


On the other hand, while their $M^\star$ are typically $>0.6~{\rm mag}$ dimmer,
their cluster finding technique might favor systems whose LF exhibits a prominent 
BCG component, resulting in a pronounced Log-normal component at the bright end, which is not observed in our sample~\footnote{
Indeed, \cite{koe07} have shown that maxBCG catalog is not biased toward bright BCGs, and bright-BCG systems do not have satellites with systematically fainter M*.}. 
We speculate that the presence of a more prominent Log-normal component anti-correlates with the LF characteristic magnitude, in the sense that 
systems with a stronger Log-normal component have dimmer satellites.

\cite{pop05} find a steeper faint end and brighter  $M^\star$ ($\alpha=-1.33\pm0.06$, $M^\star=-22.17\pm0.20$ within $1\ {\rm Mpc}\ {\rm h}^{-1}_{100}$)
than we do, but their results strongly vary depending on the
extraction region that they use. Their two-component fit within both $1.0\ {\rm Mpc}\ {\rm h}^{-1}_{100}$ and $2.0\ {\rm Mpc}\ {\rm h}^{-1}_{100}$ is in fact roughly consistent with our $\alpha$ value  ($\alpha=-1.05\pm0.13$ and $\alpha=-1.03\pm0.14$, respectively), while they 
observe a much shallower trend within $1.5\ {\rm Mpc}\ {\rm h}^{-1}_{100}$ ($\alpha=-0.76\pm0.13$).
When comparing LFs extracted within the same physical radius
R$_{200}$, the steepness of the best-fit bright Schechter component
from a later work by the same authors~\citep{pop06} is instead in
agreement with our findings ($\alpha=-1.09 \pm 0.09$, while we measure
$\alpha=-1.15 \pm 0.02$ over the whole R$_{200}$ sub-sample); their
characteristic magnitude is only slightly brighter than our measured
value ($M^\star=-21.71\pm0.16$ against $M^\star=-21.43\pm0.6$). At
the faintest magnitudes ($M_r\lse -18$)~\citet{pop06} detect a
significant upturn which is not seen in our data (see
Fig.~\ref{fig:GMA_vs_Colless}). However, we do not probe magnitudes
fainter that $M_r<-16$; also note that these authors adopted the
Colless cumulation approach which can result in a steeper LF, in
particular when the sample is weighted using only galaxies
much brighter than the cumulative LF limit ($\S$ \ref{sec:compositeLF}).

\cite{Val04} find that artificially steep faint end slopes might be caused by projection effects resulting from background galaxies, which cannot be corrected for by subtracting background fields for 2d selected clusters with no significant 3d counterpart. Despite the 2d selection algorithm applied to compile the NoSOCS catalogue, the reality of NoSOCS clusters was verified both by photometric redshifts and by X-ray analysis of a sub-sample of NoSOCS clusters \citep{Lop06}. In any case, we do not observe the steep LF faint end described by \cite{Val04}.

\begin{figure}
\includegraphics[width=0.5\textwidth,angle=0]{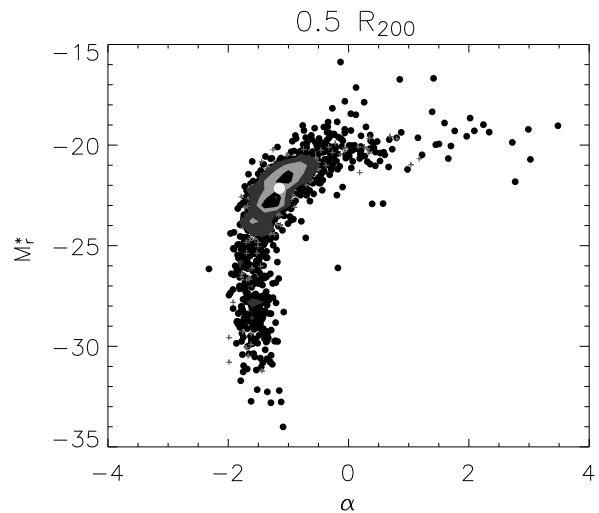}
\caption{Results of the individual fit to a single Schechter function for all
  clusters in our  sample within $0.5\ {\rm R}_{200}$. Rich clusters (Richn$_{\rm ML}\ge 9$) are plotted as gray crosses; all other clusters are shown as black dots.  Over-plotted are the contours including $25\%$, $50\%$ and  $75\%$ of the  whole distribution. A white dot shows the result of the ML best fit obtained using a single Schechter function.}
\label{fig:single_fit}
\end{figure}

In Fig.~\ref{fig:single_fit} we plot the results of the individual
best-fits to all clusters within 0.5 R$_{200}$. The individual LFs
show a wide scatter, due to the combined effects of variable S/N
levels (un-physical results are obtained for many low S/N ratio or
very bright completeness limit systems) and intrinsically different
galaxy populations (see $\S$ \ref{sec:richness}).  Despite this, the
most likely values within the whole sample are in excellent agreement
with the best fit of the ML fit obtained using a single Schechter
function (plotted as a white circle).

In the following sections we split our sample in sub-samples of
clusters spanning small ranges of richness, cluster-centric distance
and redshift in order to minimize cumulation biases and understand the
main parameters driving galaxy evolution in different environments.

\subsection{LF Dependence on the Environment}
\label{sec:richness}
Discordant results have been reported in literature in the past years,
about $M^*$ being brighter in clusters than in the field, with an
increasing trend for higher
mass-systems~\citep{DeP03,Han09,bar07,Bar09,Cra09}.  This is in
agreement with hierarchical models for galaxy formation and evolution,
where the frequency of mergers increases in intermediate and high mass
systems, such as rich groups and clusters, causing galaxies in
structures to be typically brighter than in the field, hence resulting
in a brighter M*.  

Substantial differences in the shape of the LF at different selection
radii have been found for clusters at both
low~\citep{bar07,pop06,Lob97,Rob10} and moderate-to-high
redshift~\citep{Cra09}.  While the slopes of the LFs are similar at
bright magnitudes, the main differences arise at the faint end where
the influence of the dwarf galaxies causes an increase in the
steepness of $\alpha$ with cluster-centric distance.
The sampling depth and the effective cluster-centric distance thus
have a great influence on the measured shape of the LF since the
inclusion of different fractions of the dwarf galaxy population will
directly impact the slope of the faint end.  Similar results have also
been reported for field galaxies~\citep{Xia06}. On the other hand, a
lack of significant trends with cluster-centric radius, has also been
reported~\citep{Han09} although limited to
brighter magnitudes \citep{Rud09}.
A dependence of the faint-end slope on the mass of
the cluster has also been reported, suggesting that
more massive clusters exhibit higher dwarf to giant ratios than less
massive ones~\citep{DeL04,Zan06,Gil08}.

\begin{figure*}
\centering
\includegraphics[height=0.4\textwidth,angle=-90]{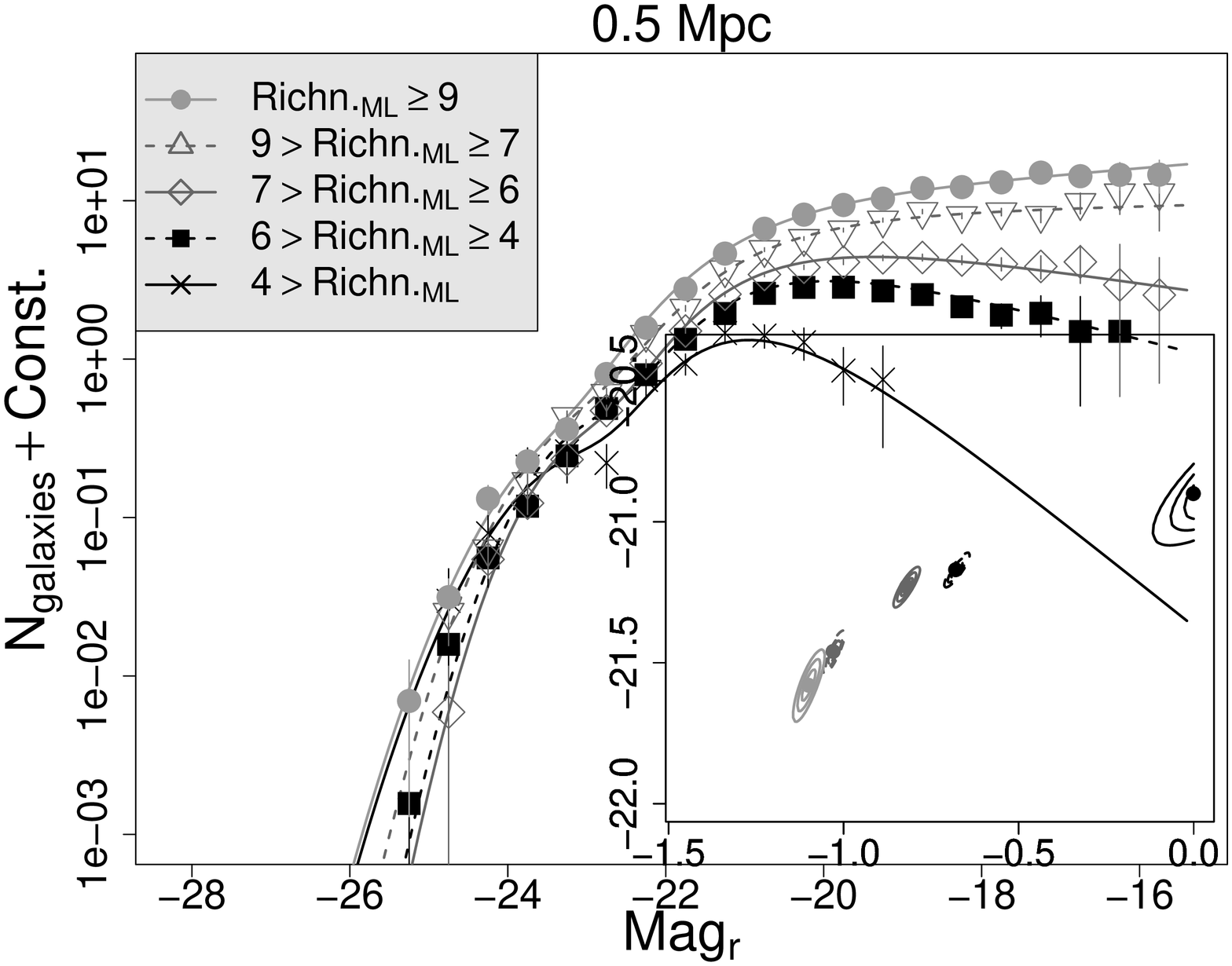}
\includegraphics[height=0.4\textwidth,angle=-90]{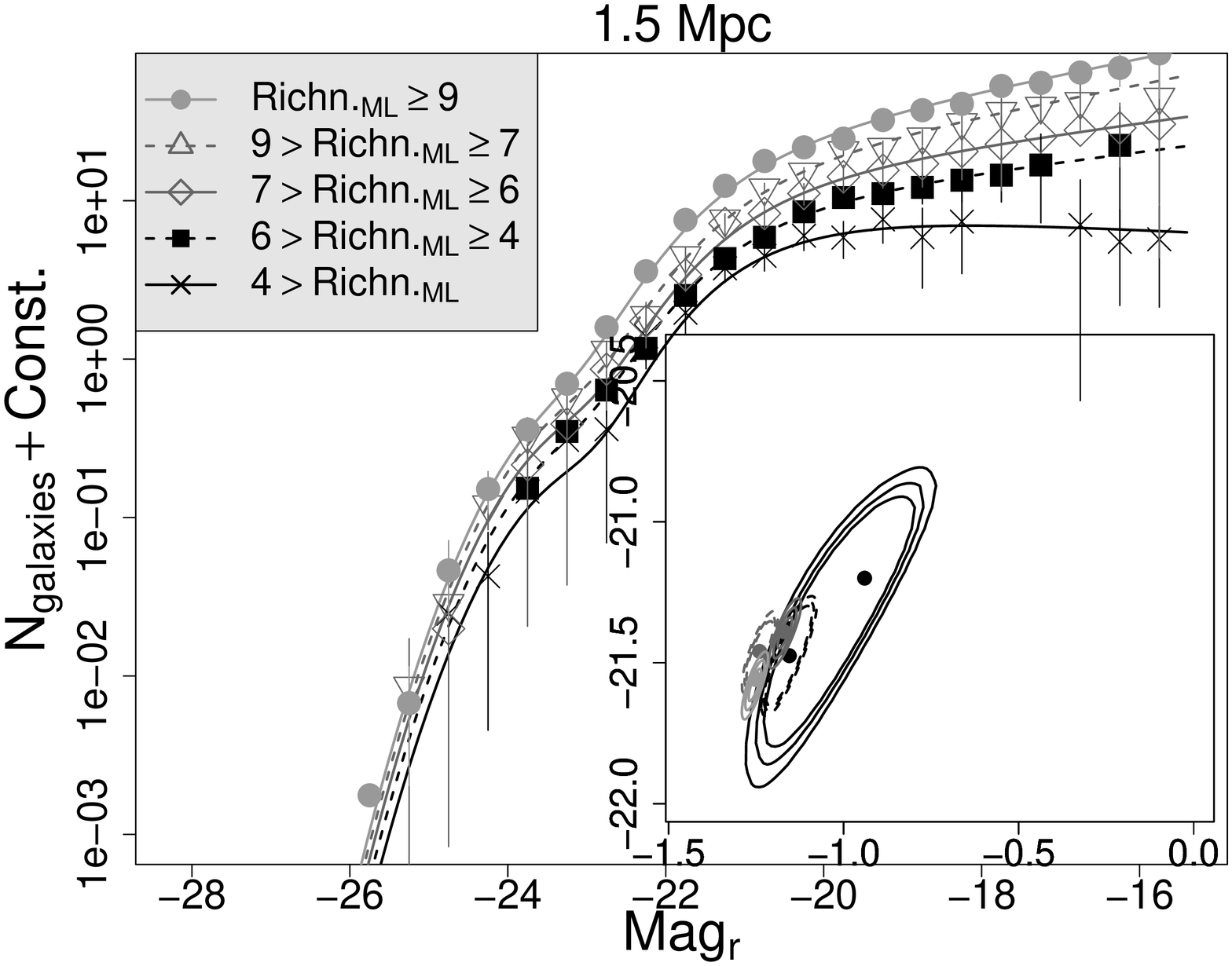}
\includegraphics[height=0.4\textwidth,angle=-90]{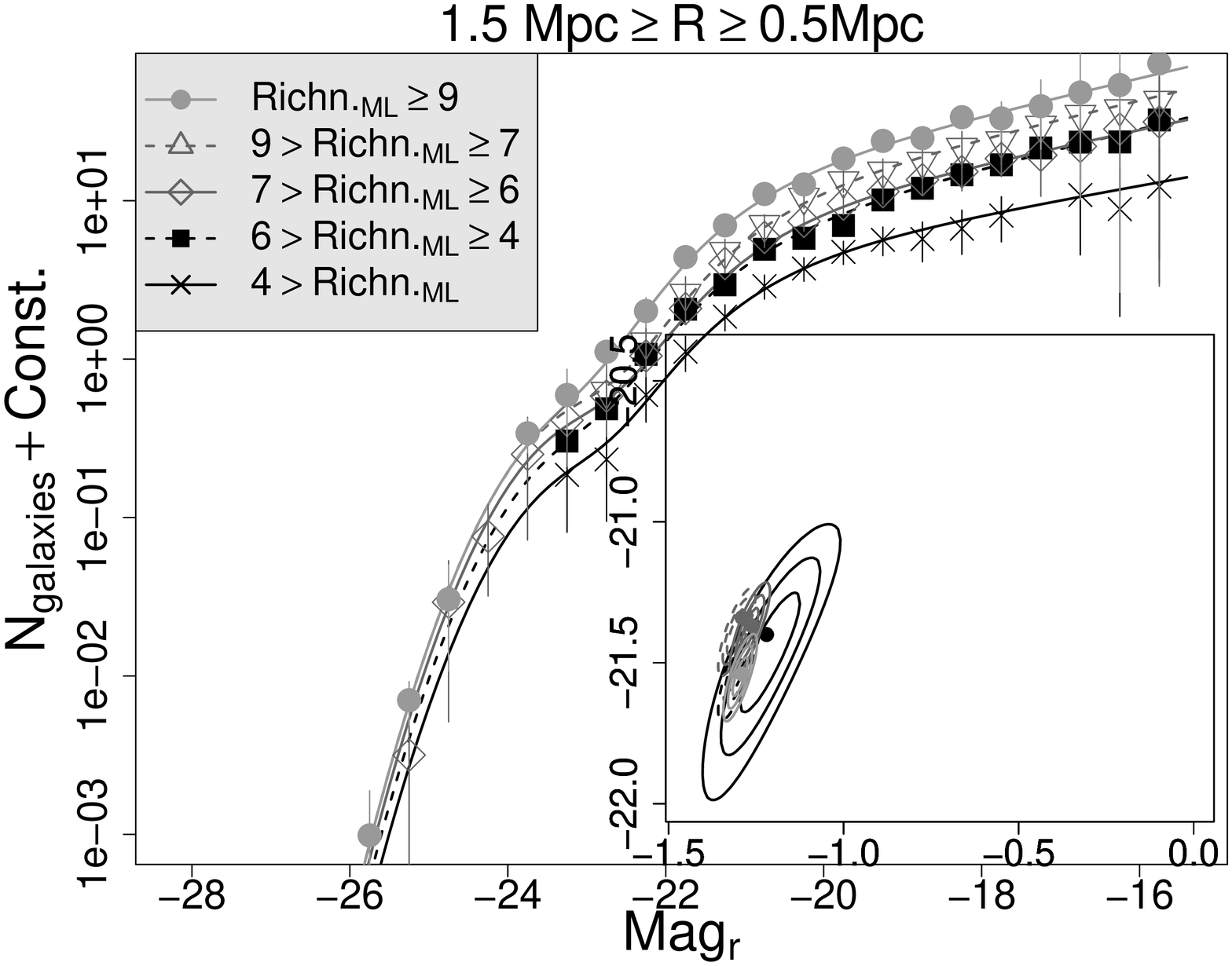}
\caption{ML fit of the LFs extracted within $0.5\ {\rm Mpc}$, $1.5\ {\rm Mpc}$, and for $0.5\ {\rm Mpc}<{\rm R}\leq 1.5$ Mpc in bins of increasing richness. Relative confidence levels are also plotted. Results are for Richn.$_{\rm ML}$ richness estimates. Data points show the composite LFs obtained using the GMA approach.}
\label{fig:LF_vs_richn_a}
\end{figure*}

\begin{figure*}
\centering
\includegraphics[height=0.4\textwidth,angle=-90]{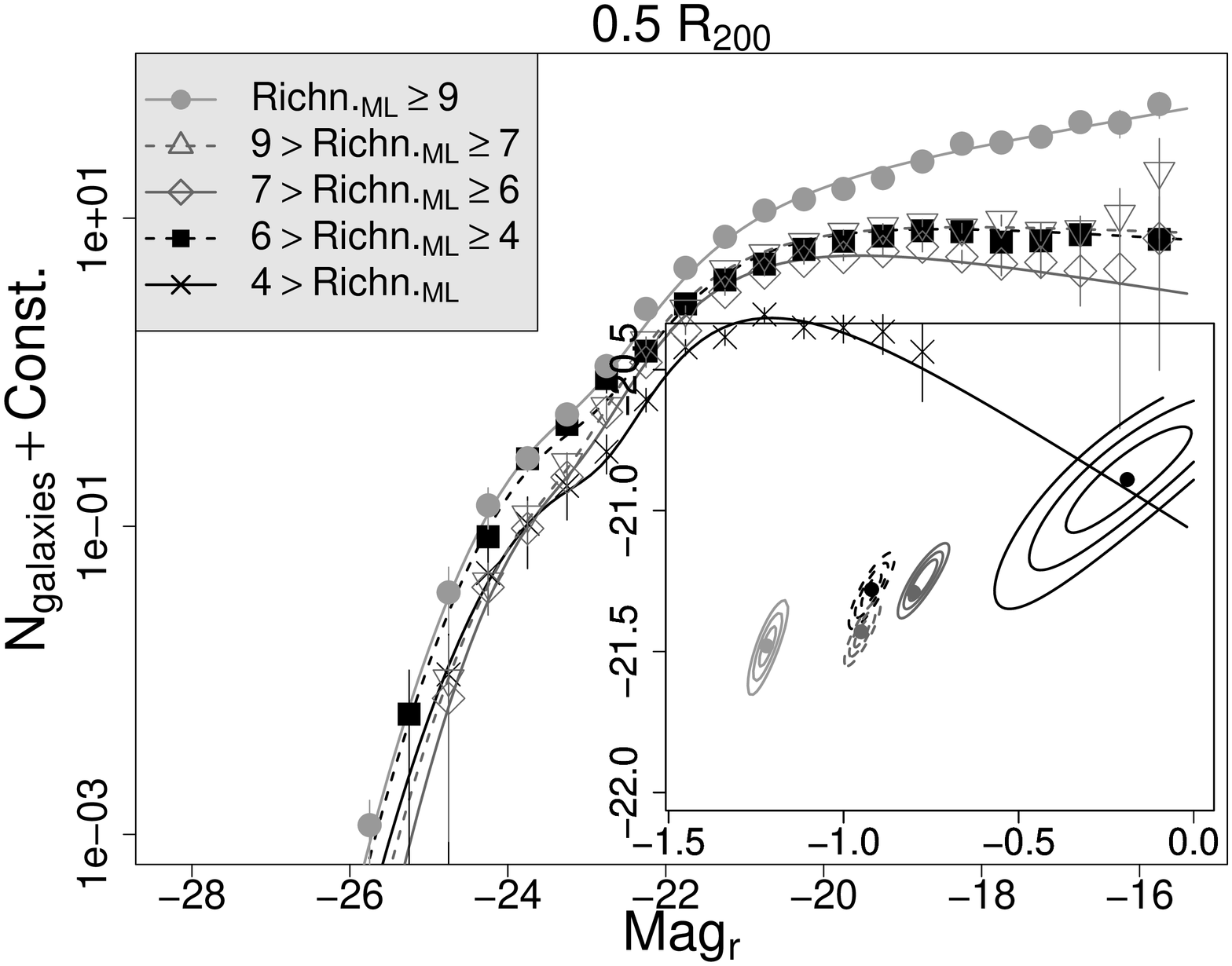}
\includegraphics[height=0.4\textwidth,angle=-90]{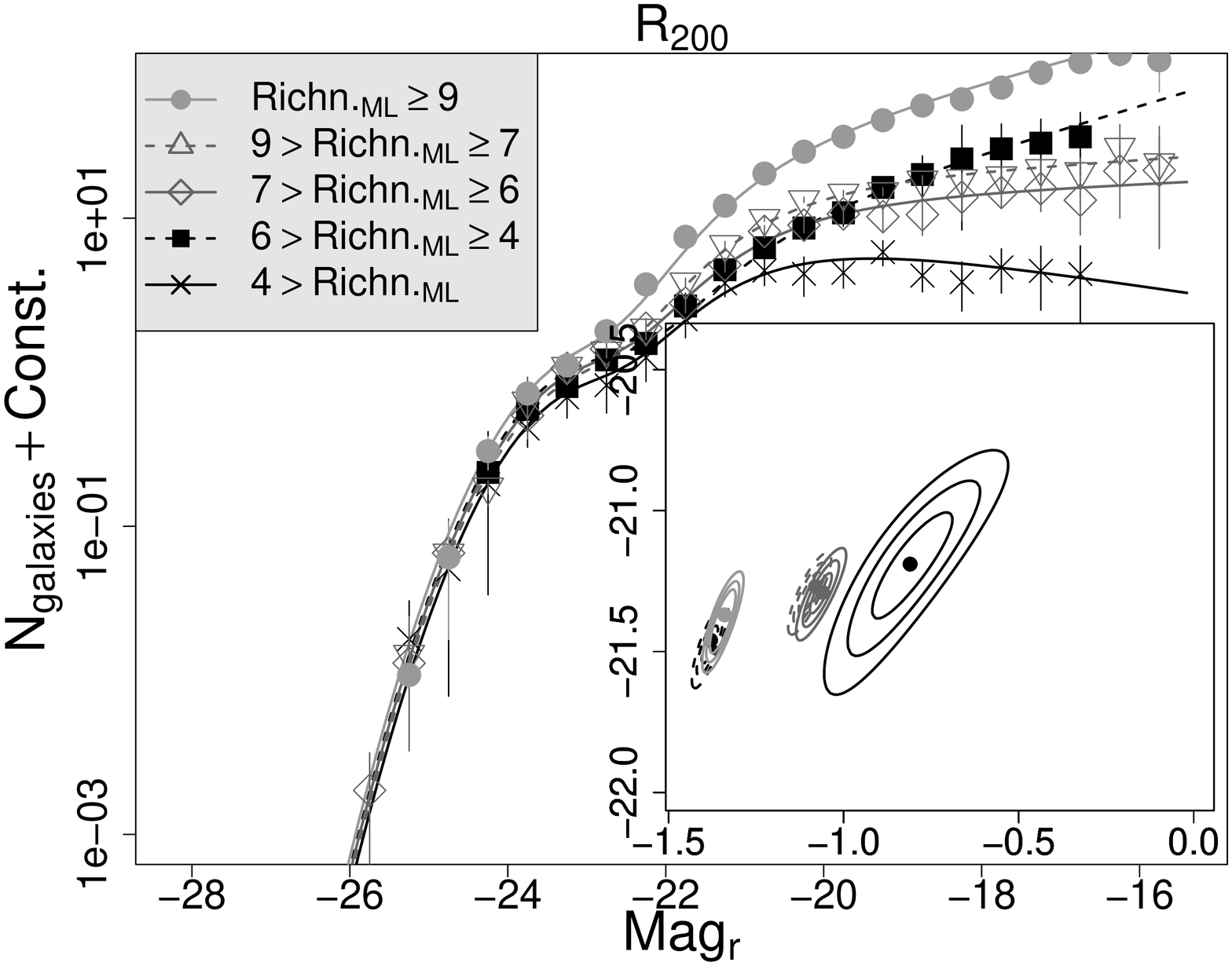}
\includegraphics[height=0.4\textwidth,angle=-90]{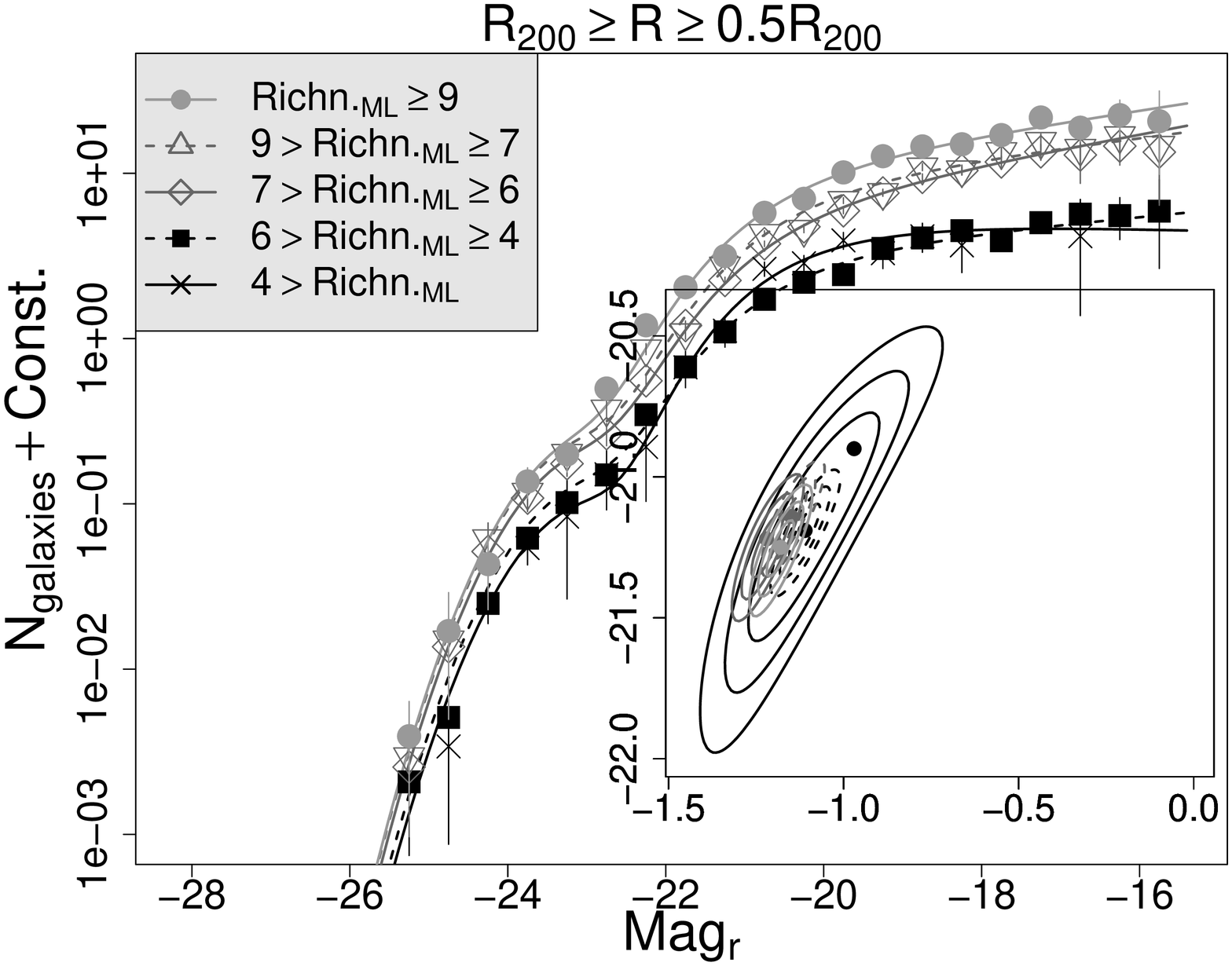}
\caption{ML fit of the LFs extracted within $0.5\ {\rm R}_{200}$, $R_{200}$ and for $0.5\ {\rm R}_{200}<{\rm R}\leq {\rm R}_{200}$ in bins of increasing richness. Relative confidence levels are also plotted. Results are for Richn.$_{\rm ML}$ richness estimates. Data points show the composite LFs obtained using the GMA approach.}
\label{fig:LF_vs_richn_b}
\end{figure*}

We start by splitting our sample into five equally populated sub-sets
of fixed richness ranges.  The top left panels of
Figs.~\ref{fig:LF_vs_richn_a} and \ref{fig:LF_vs_richn_b} show the ML
fit for the LFs in all richness bins, within a projected radius of
$0.5 {\rm Mpc}$ and $0.5 {\rm R}_{200}$, respectively. Both panels
reveal a systematic change of the overall population across the five
richness bins, in the sense that the lowest multiplicity bins show
strong dwarf suppression and fainter M$^*$. We note that the decreasing
relative normalization of the LFs is real, a result of the decreasing
cluster richness.

To explore the influence of the extraction region and of the
cluster-centric distance we fit the LFs within several extraction
radii (both fixed and physical) for each richness sub-sample: an outer
annulus ($0.5\ {\rm R}_{200}\leq{\rm R} \leq {\rm R}_{200}$ and $0.5\
{\rm Mpc}\leq{\rm R} \leq 1.5\ {\rm Mpc}$, for the physical and fixed
apertures, respectively) and a larger projected radius (${\rm R} \leq
{\rm R}_{200}$ and ${\rm R} \leq 1.5\ {\rm Mpc}$). Results of the ML
fits are listed in Table~\ref{tab:richn_bins} and shown in the bottom and 
top right panels in Figs.~\ref{fig:LF_vs_richn_a} and \ref{fig:LF_vs_richn_b}.

At larger cluster-centric radii (top right panels) we observe that the faint end
slope becomes systematically steeper for all richness bins, respect to what observed
within smaller apertures ($0.5 {\rm Mpc}$ and $0.5 {\rm R}_{200}$), while no
dramatic change is observed on the bright side. Furthermore, the
dependence on richness both of the steepness of the faint end and of
the characteristic magnitude, is very diluted.  Eventually, the bottom
panels show that in the outer annuli clusters maintain a similar
shape, within their $3 \sigma$ errors, among all richness bins. This
reveals that the sharp steepening of the faint-end and the M$^*$
brightening with richness, observed out to large projected radii, is
mainly due to the galaxies located within the central cluster regions.
Outside $\sim 0.5\ {\rm R}_{200}$
cluster galaxies appear to share similar LFs, regardless of the mass
of the parent halo mass.  It is thus likely that studies using large
apertures, either physical or fixed, are unable to identify such
trends due to the role played by the central cluster regions in
determining the LF shape.
Similar results are observed at all scales: from galaxy groups~\citep{Rob10} to rich clusters~\citep{pop06}. 

The decrease of the relative number of dwarf galaxies towards the
cluster center is most probably not due to recent merging processes, since
these are inhibited by the high velocity dispersions in cluster cores.
Also, we observe that the shape of the bright end does not strongly depend on
environment; bright early-type cannot hence be the product of cluster
environment.  The most likely explanations for these massive galaxies
is tidal collisions or collisional disruption of the dwarfs, which most
probably ends-up contributing to the intra-cluster diffuse light.

We point out that the normalization of the Log-normal component is
poorly constrained in our fits especially for poor systems, since it
is individually measured for each
cluster. 
A more detailed analysis of the bright galaxy population
is thus deferred to a future work.

\begin{table*}
\centering
 \begin{minipage}{140mm}
\caption{Results of the ML fit of the LFs extracted with different extraction radii in bins of increasing richness. Results are for Richn.$_{\rm ML}$ richness estimates.}
\label{tab:richn_bins}
\begin{tabular}{@{}lccccc@{}}
\hline 
	& Richn.$_{\rm ML}<4$& $4\leq {\rm  Richn}_{\rm ML}<6$ & $6\leq {\rm  Richn}_{\rm ML}<7$& $7\leq {\rm  Richn}_{\rm ML}<9$ & $9\leq {\rm  Richn}_{\rm ML}$\\
\hline
 \multicolumn{6}{c}{ R$\leq 0.5\ {\rm Mpc}$}\\ 
$\alpha$	&$-0.01_{-0.01}^{+0.11}$ &$-0.68_{-0.04}^{+0.04}$ &$-0.82_{-0.04}^{+0.04}$ &$-1.03_{-0.08}^{+0.07}$	&$-1.10_{-0.04}^{+0.05}$\\

M$^*$		&$-20.90_{-0.18}^{+0.12}$	&$-21.17_{-0.05}^{+0.06}$	&$-21.23_{-0.07}^{+0.07}$	&$-21.46_{-0.14}^{+0.16}$	&$-21.58_{-0.13}^{+0.13}$\\
\hline
 \multicolumn{6}{c}{ R$\leq 1.5\ {\rm Mpc}$}\\ 
$\alpha$	&$-0.9_{-0.4}^{+0.2}$	&$-1.15_{-0.07}^{+0.08}$	&$-1.17_{-0.05}^{+0.05}$	&$-1.24_{-0.05}^{+0.06}$	&$-1.25_{-0.04}^{+0.04}$\\
M$^*$		&$-21.0_{-2.0}^{+1.1}$	&$-21.47_{-0.18}^{+0.19}$	&$-21.39_{-0.14}^{+0.12}$	&$-21.46_{-0.15}^{+0.15}$	&$-21.56_{-0.14}^{+0.10}$\\
\hline
  \multicolumn{6}{c}{ $ 0.5\ {\rm Mpc}\leq$ R $\leq 1.5\ {\rm Mpc}$}\\ 
$\alpha$	&$-1.22_{-0.18}^{+0.21}$	&$-1.30_{-0.06}^{+0.06}$	&$-1.26_{-0.08}^{+0.05}$	&$-1.29_{-0.07}^{+0.04}$	&$-1.29_{-0.05}^{+0.04}$\\
M$^*$		&$-21.4_{-0.5}^{+0.4}$	&$-21.5_{-0.2}^{+0.2}$	&$-21.37_{-0.17}^{+0.15}$	&$-21.34_{-0.21}^{+0.11}$	&$-21.54_{-0.17}^{+0.14}$\\
\hline
  \multicolumn{6}{c}{ R$\leq 0.5\ {\rm R}_{200}$}\\ 
$\alpha$	&$-0.2_{-0.4}^{+0.2}$	&$-0.92_{-0.06}^{+0.07}$	&$-0.80_{-0.04}^{+0.10}$	&$-0.95_{-0.05}^{+0.06}$	&$-1.22_{-0.05}^{+0.06}$\\
M$^*$		&$-20.9_{-0.4}^{+0.3}$	&$-21.28_{-0.14}^{+0.13}$	&$-21.29_{-0.09}^{+0.18}$	&$-21.43_{-0.12}^{+0.11}$	&$-21.48_{-0.17}^{+0.16}$\\
\hline
  \multicolumn{6}{c}{ R$\leq {\rm R}_{200}$}\\ 
$\alpha$	&$-0.8_{-0.2}^{+0.3}$	&$-1.38_{-0.06}^{+0.05}$	&$-1.06_{-0.07}^{+0.07}$	&$-1.08_{-0.08}^{+0.05}$	&$-1.32_{-0.08}^{+0.04}$\\
M$^*$		&$-21.2_{-0.5}^{+0.4}$	&$-21.46_{-0.17}^{+0.13}$	&$-21.29_{-0.17}^{+0.16}$	&$-21.25_{-0.19}^{+0.10}$	&$-21.30_{-0.28}^{+0.08}$\\
\hline
  \multicolumn{6}{c}{ $ 0.5\ {\rm R}_{200}\leq$R$\leq {\rm R}_{200}$}\\ 
$\alpha$	&$-1.0_{-0.4}^{+0.3}$	&$-1.11_{-0.10}^{+0.10}$	&$-1.22_{-0.10}^{+0.08}$	&$-1.15_{-0.08}^{+0.10}$	&$-1.18_{-0.09}^{+0.09}$\\
M$^*$		&$-20.9_{-1.0}^{+0.4}$	&$-21.2_{-0.2}^{+0.2}$	&$-21.2_{-0.3}^{+0.2}$	&$-21.1_{-0.2}^{+0.2}$	&$-21.3_{-0.2}^{+0.3}$\\
\hline
\end{tabular}
\end{minipage}
\end{table*}

\subsection{Redshift Evolution}
The NoSOCS sample studied in this work spans only a limited range of
redshifts ($0.07\leq z<0.2$). In order to study the evolution of the
LF within this limited redshift range, we thus need to control the effects 
due to richness and extraction
radii discussed in the previous section, which may otherwise dominate
possible LF variations.  We hence divide our sample into rich (${\rm
  Richn.}_{\rm ML} \geq 6$) and poor (${\rm Richn.}_{\rm ML} < 6$)
clusters in order to minimize the intrinsic variation of the LF shape
(see \S~\ref{sec:richness}).  We further limit the analysis to the
central $0.5\ {\rm Mpc}$, where we have observed the strongest 
dependence of the galaxy population with the environment 
(see \S~\ref{sec:richness}).
We then analyze three redshift bins $0.07\leq z
<0.11$, $0.11\leq z <0.18$ and $0.18\leq z < 0.2$, chosen to maximize
the separation between high and low-z systems, while still retaining a
significant number of clusters in each group.

\begin{figure*}
\centering
\includegraphics[height=0.4\textwidth,angle=-90]{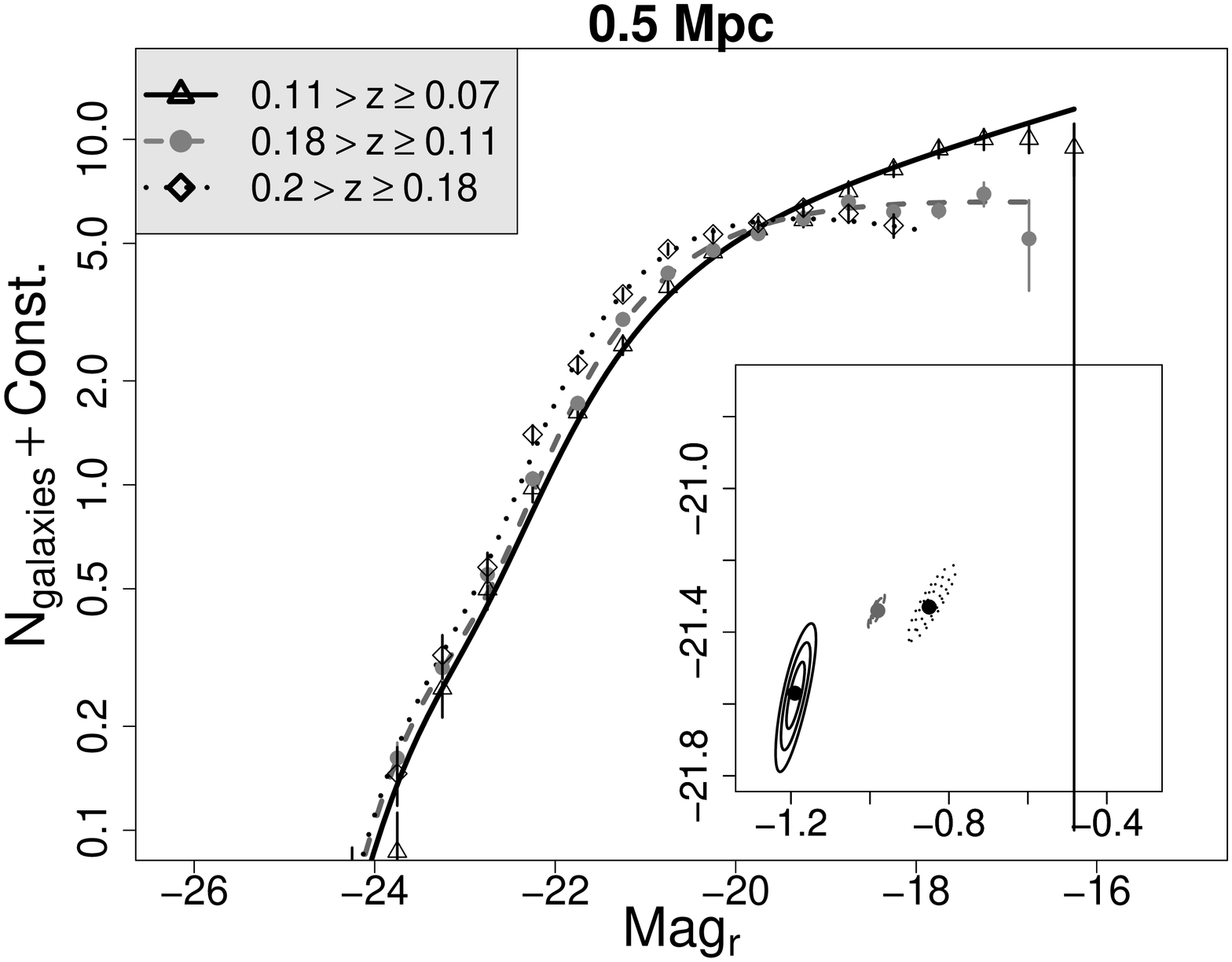}
\includegraphics[height=0.4\textwidth,angle=-90]{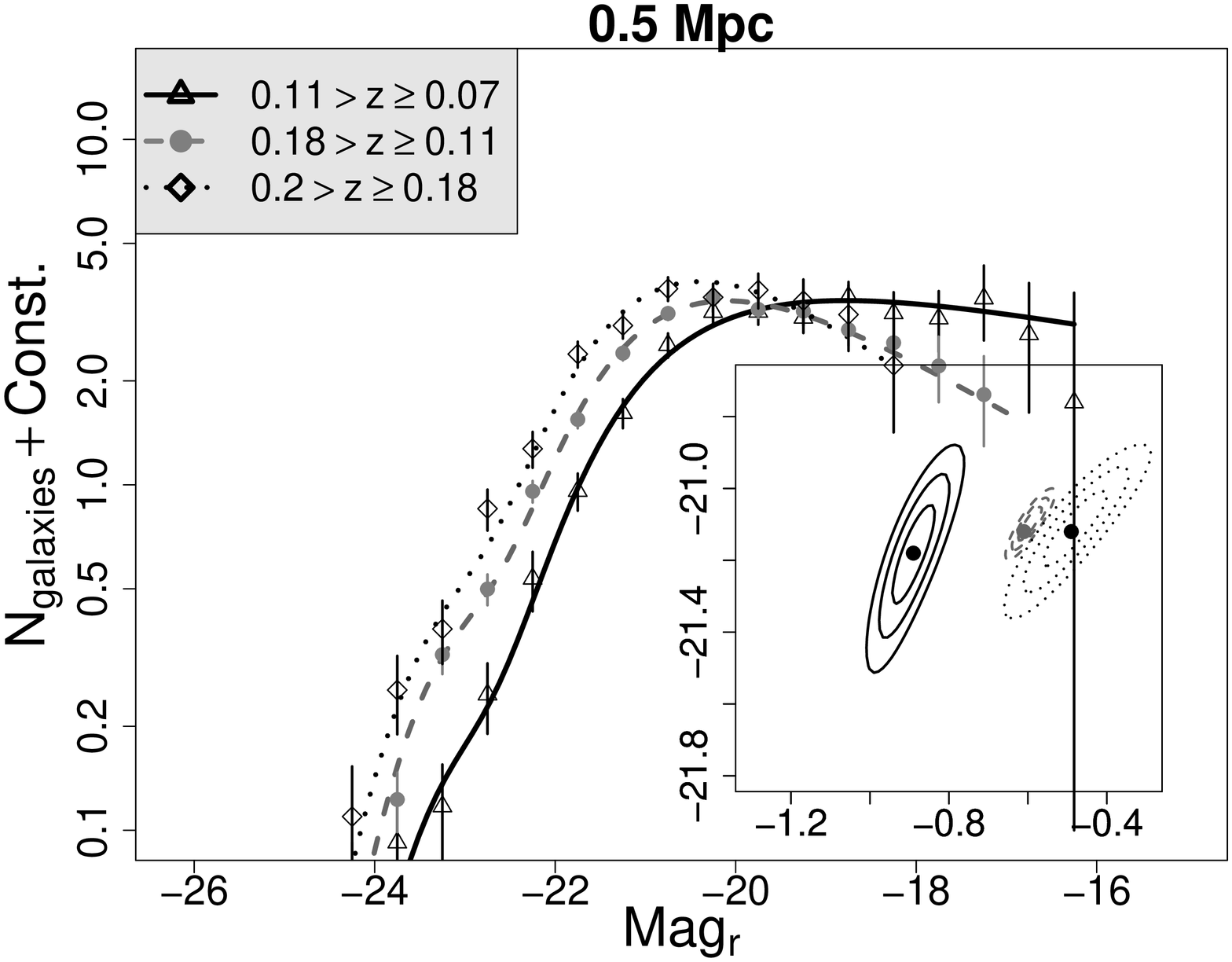}
\caption{Left panel:  ML fit of  the LFs (extracted within  $0.5\ {\rm
    Mpc}$)  of  the richest  structures  (${\rm Richn.}_{\rm  ML}>6$),
  split  into three  redshift bins,  together with  relative contours,
  corresponding to the significance levels of 1 2
  and 3~$\sigma$. Right panel: same as left panel, but for the poorest
  structures (${\rm Richn.}_{\rm ML}<6$).}
\label{fig:LF_vs_z_a}
\end{figure*}

Fig.~\ref{fig:LF_vs_z_a} 
presents the LFs derived for the three redshift bins.
We find a clear indication that the dwarf to giant ratio increases
with decreasing redshift, both for rich and poor systems.  This
confirms results (of varying significance) already reported in
literature~\citep{Kod04,Got05,Tan05,Sto07,Cra09} for cluster samples
spanning much wider redshift ranges.  Similar trends have also been
reported for field galaxies~\citep{Wil06,Xia06,Rya07}.  We underline
that the above result might be affected by the decrease in
completeness limit as a function of redshift. We measure this effect
by fitting data in the lowest redshift bin, using the completeness
limit of the highest bin. We indeed observe a decrease in the
steepness of the faint end ($\alpha=-1.06_{-0.05}^{+0.06}$ for $6\leq
{\rm Richn.}_{\rm ML}$), but still significantly different from what
measured in the highest redshift objects of our sample (see
Table~\ref{tab:z_bins_RichnML}).
It is also true that the cluster catalog completeness, as a 
function of redshift, must be taken into account in order to draw conclusions
about the LF evolution with lookback time. The completeness of the  NoSOCS catalog
has been tested through extensive mock cluster simulations, as discussed in \citet{gal09}:
the catalogue is $>80\%$ complete over the redshift range explored here for rich clusters,
while for poor systems the completeness is a strong function of redshift, dropping below 50\%
at $z>1.5$ (\citealp[see e.g. Figs. 4, 5 and 6 of][for details]{gal03}). Finding the same
redshift evolution in $\alpha$ in both rich and poor clusters thus strengthens our conclusions since the former
are less subject to completeness bias due to the NoSOCS catalog, than the latter ones.
Furthermore, any NoSOCS completeness
effect would result in a steepening of the faint end at high redshift, since the sample there would be dominated by rich systems which have a steeper faint-end slope than rich ones (Figures \ref{fig:LF_vs_richn_a} and \ref{fig:LF_vs_richn_b}). In this respect our result must be considered a conservative estimate of the LF  dependence on lookback time. 

We further measure a trend for stronger suppression of faint galaxies
(below M$^*+2$) with increasing redshift in poor systems, with respect to
more massive ones, indicating that the evolutionary stage of less
massive galaxies depends more critically on the environment.  A similar
trend has been observed for faint red galaxies by~\citet{Koy07}, while
discordant results are instead found by~\citet{Cra09,And08}.

\begin{table}
\centering
\caption{Results of  the ML fit (within a projected radius of $0.5\ {\rm Mpc}$
splitting our  sample by  redshift at
  $z=0.10$ and $z=0.17$ for both the richest ($6\leq {\rm Richn.}_{\rm
    ML}$)    and   poorest    ($4\leq   {\rm    Richn.}_{\rm   ML}<6$)
  clusters.}
\label{tab:z_bins_RichnML}
\begin{tabular}{@{}lccc@{}}
\hline
	& $0.07\leq {\rm z}<0.11$& $0.11\leq {\rm z}<0.18$ & $0.18\leq {\rm  z}<0.2$\\
\hline
\multicolumn{4}{c}{$6\leq {\rm Richn.}_{\rm ML}$ }\\ 
$\alpha$	&$-1.19_{-0.05}^{+0.04}$	&$-0.98_{-0.02}^{+0.02}$	&$-0.85_{-0.05}^{+0.07}$	\\
M$^*$		&$-21.6_{-0.2}^{+0.2}$		&$-21.34_{-0.05}^{+0.05}$	&$-21.33_{-0.09}^{+0.12}$	\\
\hline
\multicolumn{4}{c}{ $4\leq {\rm Richn.}_{\rm ML}<6$}\\ 
$\alpha$	&$-0.89_{-0.12}^{+0.13}$	&$-0.61_{-0.05}^{+0.08}$&$-0.5_{-0.2}^{+0.2}$	\\
M$^*$		&$-21.2_{-0.3}^{+0.4}$	&$-21.12_{-0.08}^{+0.12}$	&$-21.1_{-0.3}^{+0.2}$	\\
\hline
\end{tabular}
\end{table}

\section{Conclusions}
\label{sec:conclusions}
We presented the analysis of the Luminosity Function of galaxy
clusters from the Northern Sky Optical Cluster Survey, using
$r'$-band data from the Sloan Digital Sky Survey. The sample, including
1451 galaxy groups and clusters, is large enough to allow us 
to investigate in detail both the intrinsic differences in
the galaxy populations as a function of richness, cluster-centric
distance and redshift, as well as the uncertainties introduced by
different analysis techniques commonly used in the literature.

Our global LF agrees with previous studies of galaxy clusters and does
not show the presence of an ``upturn'' at faint magnitudes down to
$M_r\lse -16$, presented by some earlier works as the proof of the
presence of a very large population of dwarf galaxies
\citep[e.g.][]{pop05,Gonz06}.

We do observe a strong dependence of the LF shape ($M^\star$ and
faint-end slope) on both richness and extraction radius as expected
from the morphology-density relation and most physical models of
galaxy evolution in dense/massive systems.  The dwarf to giant ratio
increases with richness, indicating that more massive systems are more
efficient in creating/retaining a population of dwarf satellites.  Furthermore,
in the innermost regions $R\le 0.5\ {\rm R}_{200}$ we observe a
sharp steepening of the faint-end and an M$^*$ brightening with
richness.  The same effect is observed both within fixed ($0.5\ {\rm Mpc}$) 
and physical ($0.5\ {\rm R}_{200}$) apertures, suggesting that either the trend
is due to a global effect, or to a local one but operating on even
smaller scales,
and thus giving rise to similar effects.
Outside this radius, cluster galaxies appear to share
similar LFs regardless of the mass of the object.  The general trend
for the LF to become shallower with decreasing cluster-center radii
supports the hypothesis that dwarf galaxies are tidally disrupted near
the cluster center, hence providing strong evidence that the relative
mixture of giant and dwarf galaxies depends on the fraction of the
virial radius that is explored.  This also explains why some studies
using large aperture radii, even if physical, have missed these trends
in the past.  Our data further suggest that a significant growth of
this dwarf population has occurred at relatively low redshift
($z<0.2$) both in rich and poor systems.  Both the richness and radial
dependence of the faint-end slope are likely due to different mixtures
of red/passive and blue/starforming galaxy populations, as observed by
several authors (i.e.~\citealt{Zan06,bar07}), but the dwarf-to-giant
ratio of red and blue galaxies has not been investigated in detail
here and should be addressed in a forthcoming work.

We note that an appropriate richness definition is required if we want
to extract information about the environmental effects on galaxy
evolution based solely on optical data, since galaxy counts alone
spanning a large magnitude range will introduce correlations between
the LF slope and richness, which tend to dilute the observed
differences.

The results  of LF  studies such  as the one  presented here  may also
depend  strongly on  the input  cluster catalog.   \citet{gal09}
compared the NoSOCS cluster catalog to the one derived from SDSS using
the  MaxBCG algorithm  \citep{koe07}.   Due  to the  bright
magnitude limit of DPOSS, NoSOCS is an essentially flux-limited sample
with a richness-dependent completeness even at $z \sim 0.2$.  As shown
in fig.~3 of \citet{gal03}, at highest richness the percentage of
NoSOCS  recovered  clusters is  expected  to  be essentially  redshift
independent  down to  $z \sim  0.2$, while  for the  less  rich groups
analyzed in  the present study  ($N_{gal} \sim 25$; see  Sec.~2.1) the
percentage of recovery is expected to decrease by a factor of $\sim 2$
between $z \sim 0.07$ and  $z \sim 0.2$, with contamination rate being
still smaller  than $\sim 5  \%$ (see fig.~8  of \citealt{gal09}). In
contrast, the  MaxBCG method  relies on the  E/SO ridge-line  to detect
clusters, and samples such galaxies down  to $0.4 L_\star$ out to $z =
0.4$.   Thus, the  MaxBCG  catalog, trimmed  to  $z =  0.3$ to  reduce
photometric  redshift  uncertainties, provides  something  close to  a
volume-limited   sample.    Nevertheless,   the   requirement   of   a
recognizable E/S0  ridge-line may favor systems  with unevolved galaxy
populations,  resulting in  different  LF trends  than those  observed
here.   Indeed, \citep{gal09} find  that at  low  richness both
MaxBCG and  NoSOCS may likely  miss a significant fraction  of groups.
In   fact,   the   MaxBCG   catalog,  restricted   to   systems   with
$N_{gals,MaxBCG}  >  10$, misses  about  half  of  the poorest  NoSOCS
systems, despite the  $<5 \%$ contamination of NoSOCS; on the other hand,
 NoSOCS also misses many
of MaxBCG clusters with $15  > N_{gals,MaxBCG} > 10$, as expected from
the incompleteness of  NoSOCS in this low redshift  regime. This shows
how the  comparison of LFs  of low-mass systems remains  a challenging
issue,  mainly  because  of   the  different  selection  function  and
detection strategy utilized to construct different cluster catalogs.
Thus, future studies of this population will require
joining cluster catalogs generated using different algorithms to
attain a fuller picture of the true underlying population(s) and the
intrinsic LF variations. 

Finally we point out that LF studies based on large samples of
different S/N clusters and groups are extremely sensitive to the
technique used to both sum galaxies and to fit the galaxy
distributions since the LF is far from universal. The uncertainties
introduced by the different methods may explain in part the variety of
faint-end slopes reported in the literature as well as, in some cases,
the presence of a faint-end upturn. It is clear that the only way to
prevent such uncertainties would be to use data probing the same
absolute magnitude range for all clusters, such as what could be provided
by the next generation of large surveys. Otherwise extreme
care must be used in evaluating the effects that statistical
methods, in addition to the standard measurement errors, have on
the final results.

\section*{Acknowledgments}
Funding for the  Sloan Digital Sky Survey (SDSS)  and SDSS-II has been
provided  by  the  Alfred   P.  Sloan  Foundation,  the  Participating
Institutions, the National Science  Foundation, the U.S. Department of
Energy,  the  National   Aeronautics  and  Space  Administration,  the
Japanese Monbukagakusho,  and the Max  Planck Society, and  the Higher
Education  Funding   Council  for  England.  The  SDSS   Web  site  is
{\tt http://www.sdss.org/}. The authors thank Eric Feigelson e Yogesh Babu 
for their useful suggestions about the statistical treatment of the data.

\bibliographystyle{mn2e}
\bibliography{DeFilippis}

\label{lastpage}

\end{document}